\documentclass[twocolumn,aps,prb,superscriptaddress,showpacs,floatfix]{revtex4}
\usepackage{graphicx}
\usepackage{amsmath}
\usepackage{amssymb}
\usepackage{amsfonts}

\begin{document}

\title{Theoretical prediction of perfect spin filtering at interfaces between close-packed surfaces of Ni or Co and graphite or graphene}
\author{V. M. Karpan}
\affiliation{Faculty of Science and Technology and MESA$^+$ Institute
for Nanotechnology, University of Twente, P.O. Box 217, 7500 AE
Enschede, The Netherlands}
\author{P. A. Khomyakov}
\affiliation{Faculty of Science and Technology and MESA$^+$
Institute for Nanotechnology, University of Twente, P.O. Box 217,
7500 AE Enschede, The Netherlands}
\author{A. A. Starikov}
\affiliation{Faculty of Science and Technology and MESA$^+$
Institute for Nanotechnology, University of Twente, P.O. Box 217,
7500 AE Enschede, The Netherlands}
\author{G. Giovannetti}
\affiliation{Faculty of Science and Technology and MESA$^+$ Institute for Nanotechnology,
University of Twente, P.O. Box 217, 7500 AE Enschede, The Netherlands}
\affiliation{Instituut-Lorentz for Theoretical Physics, Universiteit Leiden, P. O. Box
9506, 2300 RA Leiden, The Netherlands}
\author{M. Zwierzycki}
\affiliation{Institute of Molecular Physics, P.A.N., Smoluchowskiego 17, 60-179 Pozna\'n,
Poland.}
\author{M. Talanana}
\affiliation{Faculty of Science and Technology and MESA$^+$ Institute for Nanotechnology,
University of Twente, P.O. Box 217, 7500 AE Enschede, The Netherlands}
\author{G. Brocks}
\affiliation{Faculty of Science and Technology and MESA$^+$
Institute for Nanotechnology, University of Twente, P.O. Box 217,
7500 AE Enschede, The Netherlands}
\author{J. van den Brink}
\affiliation{Instituut-Lorentz for Theoretical Physics, Universiteit Leiden, P. O. Box
9506, 2300 RA Leiden, The Netherlands}
\affiliation{Institute for Molecules and Materials, Radboud Universiteit Nijmegen, P. O.
Box 9010, 6500 GL Nijmegen, The Netherlands}
\author{P. J. Kelly }
\affiliation{Faculty of Science and Technology and MESA$^+$
Institute for Nanotechnology, University of Twente, P.O. Box 217,
7500 AE Enschede, The Netherlands}

\begin{abstract}
The in-plane lattice constants of close-packed planes of fcc and hcp Ni
and Co match that of graphite almost perfectly so that they share a
common two dimensional reciprocal space. Their electronic structures
are such that they overlap in this reciprocal space for one spin
direction only allowing us to predict perfect spin filtering for
interfaces between graphite and (111) fcc or (0001) hcp Ni or Co.
First-principles calculations of the scattering matrix show that the
spin filtering is quite insensitive to amounts of interface roughness
and disorder which drastically influence the spin-filtering properties
of conventional magnetic tunnel junctions or interfaces between
transition metals and semiconductors. When a single graphene sheet is
adsorbed on these open $d$-shell transition metal surfaces, its
characteristic electronic structure, with topological singularities at
the K points in the two dimensional Brillouin zone, is destroyed by the
chemical bonding. Because graphene bonds only weakly to Cu which has no
states at the Fermi energy at the K point for either spin, the
electronic structure of graphene can be restored by dusting Ni or Co
with one or a few monolayers of Cu while still preserving the ideal
spin injection property.
\end{abstract}


\maketitle

\section{Introduction}
\label{sec:Gr_TMR_intro} We recently predicted a perfect spin filtering
effect for ultra-thin films of graphite sandwiched between two
ferromagnetic leads. \cite{Karpan:prl07} This prediction emerged from
two rapidly developing branches of condensed matter physics:
magnetoelectronics \cite{Zutic:rmp04} and graphene electronics.
\cite{Neto:rmp08} Magneto-electronics exploits the additional degree of
freedom presented by the intrinsic spin and associated magnetic moment
of electrons while graphene electronics is based upon the unique
electronic properties of two-dimensional graphene sheets. Based on the
giant magnetoresistance effect discovered twenty years ago,
\cite{Baibich:prl88,Binasch:prb89} magnetoelectronics was rapidly
applied to making improved read head sensors for hard disk recording
and is a promising technology for a new type of magnetic storage
device, a magnetic random access memory. The giant magnetoresistance
(GMR) effect is based on the spin dependence of the transmission
through interfaces between normal and ferromagnetic metals (FM). The
effect is largest when the current passes through each interface in a
so-called current-perpendicular-to-the-plane (CPP) measuring
configuration but the absolute resistance of metallic junctions is too
small for practical applications and the current-in-plane (CIP)
configuration with a much smaller MR is what is used in practice.
Replacing the non-magnetic metal spacer with a semiconductor
\cite{Julliere:pla75} or insulator (I), such as Al$_2$O$_3$
\cite{Moodera:prl95,Miyazaki:jmmm95} results in spin-dependent
tunneling and much larger resistances are obtained with FM$|$I$|$FM
magnetic tunnel junctions (MTJs). Substantial progress has been made
in increasing the tunneling MR effect by replacing the amorphous
Al$_2$O$_3$ insulator with crystalline MgO.
\cite{Yuasa:natm04,Parkin:natm04} Though there is a relatively large
lattice mismatch of 3.8\% between Fe and MgO, the tunneling
magnetoresistance (TMR) in Fe$|$MgO$|$Fe junctions has been reported to
reach values as high as 180\% at room temperature. \cite{Yuasa:apl05a}
Low temperature values as high as 1010\% have been reported for
FeCoB$|$MgO$|$FeCoB MTJs. \cite{Lee:apl07,Yuasa:jpd07} The sensitivity
of TMR (and spin injection) to details of interface structure
\cite{Xu:prb06,Zwierzycki:prb03} makes it difficult to close the
quantitative gap between theory and experiment so it is important for
our understanding of TMR to be able to prepare interfaces where
disorder does not dominate the spin filtering properties. This remains
a challenge due to the high reactivity of the open-shell transition
metal (TM) ferromagnets Fe, Co, and Ni with typical semiconductors and
insulators.

\begin{table}[b]
\begin{ruledtabular}
\caption[Tab1]{ Lattice constants of Co, Ni, Cu, and graphite, $a_{\rm
hex} \equiv a_{\rm fcc} / \sqrt{2}$. Equilibrium separation $d_0$ for a
single graphene sheet on top of the graphite (0001) and Co, Ni or Cu
fcc (111) surfaces as calculated within the framework of the DFT-LDA
using the in-plane lattice constant $a_{\rm hex}=2.46$ \AA.}
\begin{tabular}{lllll}
                               & Graphite & Co    & Ni      & Cu     \\
\hline $a_{\rm fcc}^{\rm expt}$ (\AA) &          &
3.544\footnotemark[1]
                                                  & 3.524\footnotemark[1]
                                                            & 3.615\footnotemark[1]   \\
$a_{\rm hex}^{\rm expt}$ (\AA) & 2.46     & 2.506 & 2.492   & 2.556  \\
$a_{\rm hex}^{\rm LDA}$  (\AA) & 2.45     & 2.42  & 2.42    & 2.49   \\
${d_0}$ (\AA)                  & 3.32     & 2.04  & 2.03    & 3.18   \\
\end{tabular}
\label{tab:Gr_TMR_tableone}
\end{ruledtabular}
\footnotetext[1]{Ref.\onlinecite{Ibach:95}}
\end{table}

\begin{figure*}[btp]
\includegraphics[scale=0.7]{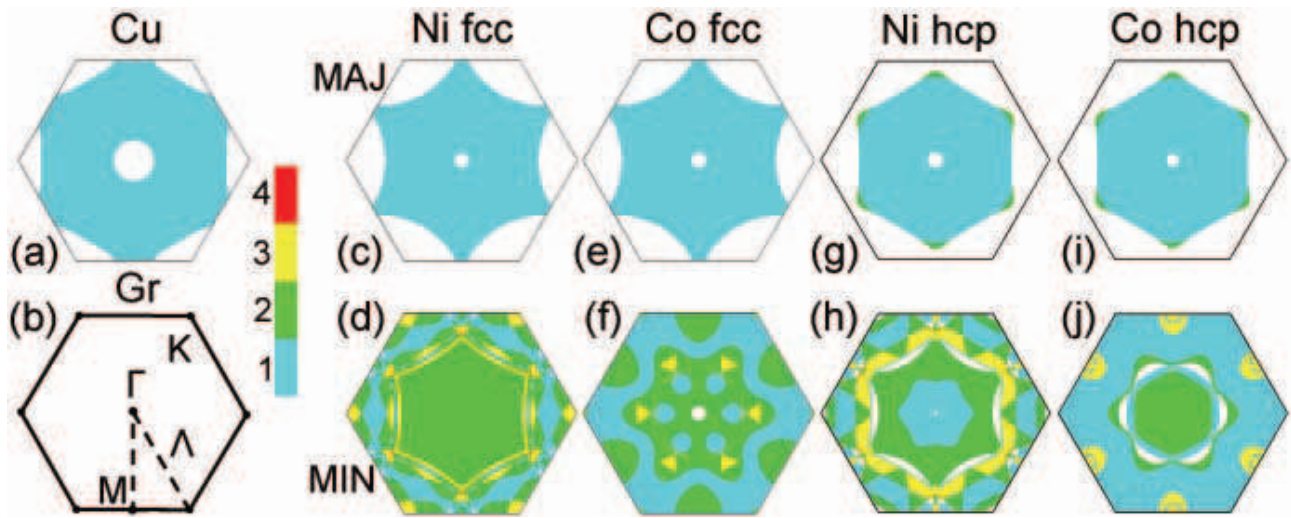}
\caption{Fermi surface projections onto close-packed planes for: (a)
fcc Cu; (c) majority- and (d) minority-spin fcc Ni (111); (e) majority-
and (f) minority-spin fcc Co (111); (g) majority- and (h) minority-spin
hcp Ni (0001); (i) majority- and (j) minority-spin hcp Co (0001). For
graphene and graphite, surfaces of constant energy are centred on the K
point of the two dimensional interface Brillouin zone (b). The number
of Fermi surface sheets is given by the colour bar.}
\label{fig:Gr_TMR_fig5}
\end{figure*}

With this in mind, we wish to draw attention to a quite different
material system in which a thin graphite film is sandwiched between two
ferromagnetic leads. Graphite is the ground state of carbon and as one
of the most important elemental materials, its electronic structure has
been studied in considerable detail. It consists of weakly interacting
sheets of carbon atoms strongly bonded in a very characteristic
honeycomb structure. Because of the weak interaction between these
``graphene'' or ``monolayer graphite'' sheets, the electronic structure
of graphite is usually discussed in two steps: first, in terms of the
electronic structure of a single $sp^2$-bonded sheet, followed by
consideration of the interaction between sheets.
\cite{Wallace:pr47,Lomer:prsla55,Slonczewski:pr58} From these early,
and many subsequent studies, it is known that graphene is a ``zero-gap
semiconductor'' or a semimetal in which the Fermi surface is a point at
the ``K'' point in the two-dimensional reciprocal space. The physical
properties associated with this peculiar electronic structure have been
studied theoretically in considerable detail, in particular in the
context of carbon nanotubes which can be considered as rolled-up
graphene sheets. \cite{Ando:jpsj05} With the very recent discovery and
development of an exceptionally simple procedure for preparing single
and multiple graphene sheets, micromechanical cleavage,
\cite{Novoselov:pnas05} it has became possible to probe these
predictions experimentally. Single sheets of graphene turn out to have
a very high mobility \cite{Novoselov:sc04} that manifests itself in a
variety of spectacular transport phenomena such as a minimum
conductivity, anomalous quantum Hall effect (QHE),
\cite{Novoselov:nat05,Zhang:nat05} bipolar supercurrent
\cite{Heersche:nat07} and room-temperature QHE.\cite{Novoselov:sc07}
Spin injection into graphene using ferromagnetic electrodes has already
been realized. \cite{Hill:ieeem06,Tombros:nat07} The weak spin-orbit
interaction implied by the low atomic number of carbon should translate
into very long intrinsic spin-flip scattering lengths, a very desirable
property in the field of spin electronics or ``spintronics'', which
aims to combine traditional semiconductor-based electronics with
control over spin degrees of freedom. However, the room temperature
two-terminal MR effect of $\sim 10\%$ observed in lateral,
current-in-plane (CIP) graphene-based devices with soft permalloy leads
is still rather small. \cite{Hill:ieeem06}

Instead of a CIP geometry, we consider a CPP TM$|$Gr$|$TM (111)
junction, where TM is a close-packed surface of fcc or hcp Ni or Co and
Gr is graphite (or $n$ sheets of graphene, Gr$_n$). We argue that such
a junction should work as a perfect spin filter. The essence of the
argument is given by Table~\ref{tab:Gr_TMR_tableone} and
Fig.~\ref{fig:Gr_TMR_fig5}. According to
Table~\ref{tab:Gr_TMR_tableone}, the surface lattice constants of (111)
Ni, Co and Cu match the in-plane lattice constants of graphene and
graphite almost perfectly. The lattice mismatch of 1.3\% at the
Ni(111)$|$Gr interface is, in fact, one of the smallest for the
magnetic junctions that have been studied so far. This small lattice
mismatch suggests that epitaxial TM$|$Gr$|$TM junctions might be
realized experimentally, for example using chemical vapor deposition.
\cite{Dedkov:apl08,Oshima:jpcm97,Gamo:ss97} Assuming perfect lattice
matching at the TM$|$Gr interface, it is possible to directly compare
the Fermi surface projection of graphite with the projections of the
Fermi surfaces (FS) of fcc Cu and of fcc and hcp Ni and Co onto
close-packed planes, see Fig.~\ref{fig:Gr_TMR_fig5}.

The Fermi surface of graphene is a point at the high-symmetry K point
in reciprocal space. The Fermi surfaces of graphite and of doped
graphene are centred on this point and close to it.
Figure~\ref{fig:Gr_TMR_fig5} shows that there are no majority spin
states for Ni and Co close to the K point whereas minority spin states
exist (almost) everywhere in the surface BZ. Only the minority spin
channel should then contribute to transmission from a close-packed TM
surface into graphite. In a TM$|$Gr$|$TM junction, electrons in other
regions of reciprocal space on the left electrode would have to tunnel
through graphite to reach the right electrode.  If the graphite film is
taken thick enough to suppress tunneling, majority spin conductance
will be quenched and only minority spin conductance through the
graphite will survive {\em i.e.} perfect spin filtering will occur
when the magnetizations are aligned in parallel (P). For antiparallel
(AP) alignment, the conductance will vanish.

In this paper, we wish to study the effectiveness of this spin
filtering quantitatively: how it depends on the thickness of the
graphite film, the geometry of the clean metal-graphite interface,
interface roughness and disorder, and lattice mismatch. While we will
be mainly concerned with the CPP geometry, we will also comment on the
applicability of some of our conclusions to the CIP geometry. The paper
is organized as follows. In Sec.~\ref{sec:Gr_TMR_method} we give a
brief description of the computational method and outline the most
important technical details of the calculations. The transport
formalism we use is based upon a very efficient minimal basis of
tight-binding muffin tin orbitals (TB-MTO) in combination with the
atomic spheres approximation (ASA).\cite{Andersen:85} While the ASA
works well for close-packed structures, some care is needed in using it
for very open structures like that of graphite. In
Sec.~\ref{sec:Gr_TMR_geometry} we therefore benchmark the electronic
structures calculated using the TB-MTO-ASA method with those obtained
from plane-wave pseudopotential calculations.
Section~\ref{sec:Gr_TMR_transport} contains the results of
spin-dependent electron transport calculations for specular interfaces
(ideal junction) as well as for junctions with interface roughness and
alloy disorder. A summary is given and some conclusions drawn in
Sec.~\ref{sec:Gr_TMR_conclusions}.

\section{Computational Method}
\label{sec:Gr_TMR_method}

The starting point for our study is an atomic structure calculated by
minimizing the total energy within the local spin density approximation
(LSDA) of density functional theory (DFT). This was done using a
plane-wave pseudopotential (PWP) method based upon projector augmented
wave (PAW) pseudopotentials \cite{Blochl:prb94b} as implemented in the
VASP program.\cite{Kresse:prb99,Kresse:prb93,Kresse:prb96} The
interaction between graphite and the TM surface is modelled using a
repeated slab geometry of six metal layers with a graphene sheet on top
and a vacuum thickness of $\sim 12$ \AA. To avoid interactions between
periodic images of the slab, a dipole correction is applied.
\cite{Neugebauer:prb92} The surface Brillouin zone (SBZ) was sampled
with a $36 \times 36$ $\mathbf{k}$-point grid and the SBZ integrals
carried out with the tetrahedron integration scheme.
\cite{Blochl:prb94a} A plane wave kinetic energy cutoff of 400 eV was
used. The plane-wave pseudopotential calculations yield energy band
structures, charge transfers, binding energies and work functions for
single TM$|$Gr interfaces. \cite{Karpan:prl07,Giovannetti:prl08} The
equilibrium distances $d_0$ between the graphene sheet and the TM
surfaces are summarized in Table~\ref{tab:Gr_TMR_tableone}.

The equilibrium geometries are used as input for self-consistent TB
linearized MTO (TB-LMTO) \cite{Andersen:85} calculations for the
TM$|$Gr$_n|$TM junction. The resulting Kohn-Sham potentials are used to
calculate spin-dependent transmission probabilities through the
TM$|$Gr$_n|$TM junction using a TB-MTO wave-function matching
\cite{Ando:prb91,Khomyakov:prb05} scheme.
\cite{Xia:prb01,Xia:prb06,Zwierzycki:pssb08} To do this, the junction
is divided into three parts consisting of a scattering region
sandwiched between semi-infinite left and right leads, all of which are
divided into layers that are periodic in the lateral direction. The
leads are assumed to be ideal periodic crystals in which the electron
states (modes) are wave functions with Bloch translational symmetry. By
making use of its Bloch symmetry, a semi-infinite lead can be
represented as an energy-dependent non-Hermitian potential on the
boundary of the scattering region so that the infinite system is made
finite. According to the Landauer-B\"uttiker formalism of transport,
the conductance can be calculated by summing up all the probabilities
for transmitting an electron from the electron modes in the left lead
through the junction into electron modes in the right leads.
\cite{Buttiker:prb85,Datta:95,Xia:prb06}

The effect of various types of disorder on the transmission can be
studied using the same formalism and computer codes by modelling the
disorder within large lateral supercells \cite{Xia:prb01,Xia:prb06} and
averaging over many configurations of disorder generated by choosing
positions of impurity atoms or imperfections randomly. We study three
types of disorder: interface roughness, interface alloying and lattice
mismatch. In the first two cases, averaging is performed over a minimum
of ten configurations of disorder. To model interface roughness, some
surface atoms are removed (replaced by ``empty spheres'' with nuclear
charges that are zero in the ASA) and the ASA potentials are calculated
self-consistently using a layer version \cite{Turek:97} of the coherent
potential approximation (CPA). \cite{Soven:pr67} The effect of
interface alloying which might occur if deposition of a thin layer of
Cu on Ni or Co (``dusting'') leads to intermixing is modelled in a
similar fashion. Thirdly, the small lattice mismatch between graphite
and TM is modelled by ``cutting and pasting'' AS potentials from
self-consistent calculations for TM$|$Gr$_n|$TM junctions with two
different in-plane lattice constants. The two systems are then combined
using a supercell whose size is determined by the lattice mismatch. For
self-consistent TB-LMTO-ASA calculations, the BZ of lateral supercells
is sampled with a density roughly corresponding to a $24 \times 24$
$\mathbf{k}$-point grid for a $1\times1$ interface unit cell. To
converge the conductance, denser grids containing $800 \times 800$, $20
\times 20$ and $8 \times 8$ $\mathbf{k}$-points are used for $1 \times
1$ (ideal junction), $5 \times 5$ and $20 \times 20$ lateral
supercells, respectively.

\section{Geometry and electronic structure of $\rm TM|Gr_n|TM$}
\label{sec:Gr_TMR_geometry}

\begin{figure}[btp]
\includegraphics[scale=0.45]{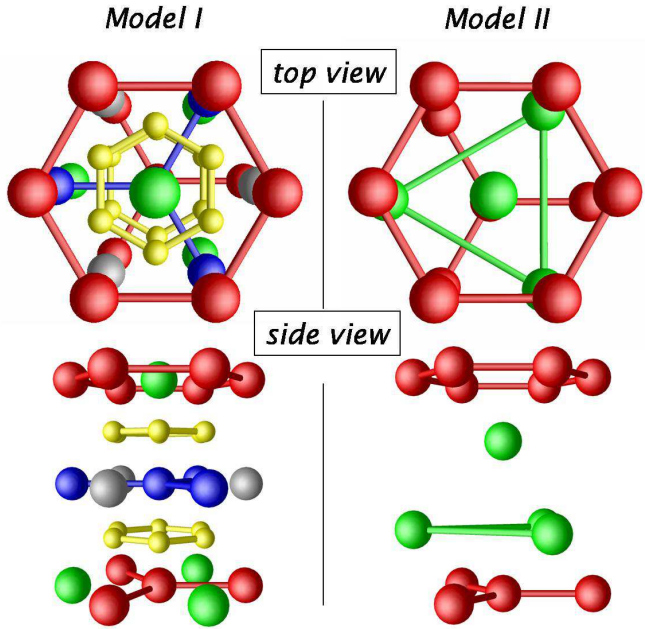}
\caption{Top and side perspective views (top and bottom panels) of
graphite where the potential is represented in the atomic spheres
approximation using additional, empty atomic spheres. Model I (left)
contains 32 empty spheres in a unit cell containing 4 carbon atoms (red
spheres). Model II (right), contains just 4 empty spheres. For model I,
gray, green, blue and yellow spheres display the positions of the empty
spheres E$_1$, E$_2$, E$_3$ and E$_4$, respectively. For model II,
there is just one type of empty sphere (green).}
\label{fig:Gr_TMR_fig1}
\end{figure}

In this section we describe in more detail how the electronic structure
of TM$|$Gr$_n|$TM junctions for TM=Cu, Ni or Co is calculated. These
close-packed metals can be grown with ABC stacking in the (111)
direction (fcc), or with AB stacking in the (0001) direction (hcp). We
neglect the small lattice mismatch of 1.3\%, 1.9\% and 3.9\% for the
Ni$|$Gr, Co$|$Gr, and Cu$|$Gr interfaces, respectively, and assume the
junction in-plane lattice constant to be equal to that of graphite,
$a_{\rm Gr}=2.46$ \AA. In the atomic spheres approximation, the atomic
sphere radii of Ni, Co and Cu are then $r_{\rm TM}=2.574$ a.u. The ASA
works well for transition metals like Co, Ni or Cu which have
close-packed structures. For materials like graphite which has a very
open structure with an in-plane lattice constant $a_{\rm Gr}=2.46$ \AA,
and an out-of-plane lattice constant $c_{\mathrm{Gr}}=6.7$ \AA, the
unmodified ASA is not sufficient. Fortunately, a reasonable description
of the crystal potential can be obtained by packing the interstitial
space with empty spheres. \cite{Glotzel:ssc80} This procedure should
satisfy the following criteria: (i) the total volume of all atomic
spheres has to be equal to the volume of the entire system (space
filling), and the (ii) overlap between the atomic spheres should be as
small as possible.

\begin{table}[btp]
\begin{ruledtabular}
\caption[Tab2]{Wyckoff symbols, standardized position parameters and
atomic sphere radii for carbon atoms, C, and empty spheres, E (with
nuclear charge Z=0), for two structural models of graphite with space
group $D_{6h}^4$~(P6$_3$/mmc)~Ref.\onlinecite{ITC}. Model I contains
four different types of empty sphere: E$_1$, E$_2$, E$_3$, E$_4$; model
II only one, E. }
\begin{tabular}{lllll}
Model & Atom  &  Wyckoff  &  position             & radius  \\
      &       & position  & parameters            & (a.u.)  \\
\hline
  I   & C$_1$ &    2b     &                       & 1.56    \\
      & C$_2$ &    2c     &                       & 1.56    \\
      & E$_1$ &    2a     &                       & 1.4     \\
      & E$_2$ &    2d     &                       & 1.6     \\
      & E$_3$ &    4f     &  z=0.5                & 1.4     \\
      & E$_4$ &   24l     &  x=1/3, y=0, z=0.38   & 0.9     \\
\hline
  II  & C$_1$ &    2b     &                       & 1.56    \\
      & C$_2$ &    2c     &                       & 1.56    \\
      & E     &    4f     &  z=0.4                & 2.18    \\
\end{tabular}
\label{tab:Gr_TMR_tabletwo}
\end{ruledtabular}
\end{table}

\subsection{Graphite and graphene}

\begin{figure}[!b]
\includegraphics[scale=0.65]{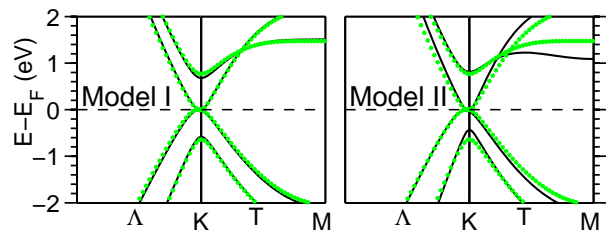}
\caption{(Color online) Band structure of graphite for model I (on the left),
and model II (on the right). Gray (green) dots and black lines correspond to
band structures calculated using the PWP and TB-MTO-ASA methods, respectively.}
\label{fig:Gr_TMR_fig2}
\end{figure}

To see how this procedure works in practice, we benchmark the
TB-MTO-ASA band structure of graphite against the ``exact'' band
structure calculated with the PWP method. To preserve the graphite
$D_{6h}^4$~(P6$_3$/mmc) space group symmetry, \cite{ITC} the positions
of the atomic spheres are chosen at Wyckoff positions. There are twelve
different Wyckoff positions consistent with $D_{6h}^4$ symmetry and the
best choice of empty spheres is not immediately obvious. We construct
two models that describe the band structure close to the Fermi energy
well compared to the PWP results; this is what is most relevant for
studying transport in the linear response regime. Model I with 32 empty
spheres per unit cell and model II with only 4 empty spheres per unit
cell both preserve the symmetry of graphite within the ASA. The crystal
structures of graphite packed with empty spheres according to these two
models is shown schematically in Fig.~\ref{fig:Gr_TMR_fig1}. Note that
not all the empty spheres in a unit cell are shown in the figure. The
Wyckoff labels, atomic sphere coordinates and radii are given in
Table~\ref{tab:Gr_TMR_tabletwo}. Figure~\ref{fig:Gr_TMR_fig2} shows the
band structure of graphite obtained with the TB-MTO-ASA for models I
and II compared to the ``exact'' PWP band structure. Both models are
seen to describe the graphite $\pi$ bands around the Fermi energy very
well. Model I provides a very good description of the bands within $\pm
2$~eV of the Fermi energy, while the smaller basis model II is quite
good within $\pm 1$~eV. At the cost of including many more empty
spheres, model I provides a better description of the crystal potential
between the graphene planes than model II. For this reason we use model
I to study the transport properties of ideal junctions, junctions with
interface roughness and alloy disorder. To be able to handle the large
$20 \times 20$ lateral supercells needed to model a lattice mismatch of
5\% at the TM$|$Gr interface, we use model II.

\subsection{Graphene on Ni(111) substrate}

\begin{figure}[!b]
\includegraphics[scale=0.34]{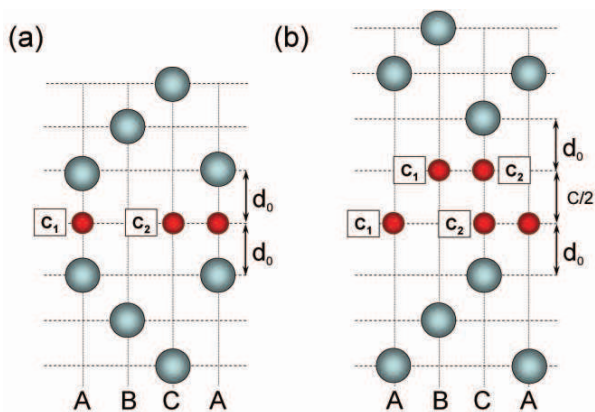}
\caption{(Color online) ``AC'' model of TM$|$Gr$_n|$TM structure for (a) odd
and (b) even numbers of graphene sheets. Carbon atoms are represented by
small dark (red) spheres, TM atoms by larger gray spheres. The configuration
shown in (a) is a c$_1$c$_1$ configuration with the carbon atom labelled
c$_1$ above an ``A'' site surface layer TM atom of the top and the
bottom electrodes. The other carbon atom, c$_2$, is above a third layer
TM atom on a ``C'' site. An equivalent c$_2$c$_2$ configuration in
which the c$_2$ atoms are on top of ``A'' site TM atoms can be realized
by rotating the top and bottom electrodes by $180^{\circ}$ about a
vertical axis through the second layer ``B'' sites; this effectively
interchanges c$_1$ and c$_2$. Two other equivalent configurations
c$_1$c$_2$ and c$_2$c$_1$ can be realized in an analogous fashion by
rotating either the top or the bottom electrode through $180^{\circ}$.
For two sheets of graphene stacked as in graphite, a c$_2$c$_2$
configuration is sketched in (b). Interlayer distance is indicated as
$d_0$ and $c/2$ is the distance separating two neighbouring graphene
sheets.} \label{fig:Gr_TMR_fig3}
\end{figure}

\begin{figure}[btp]
\includegraphics[scale=0.65]{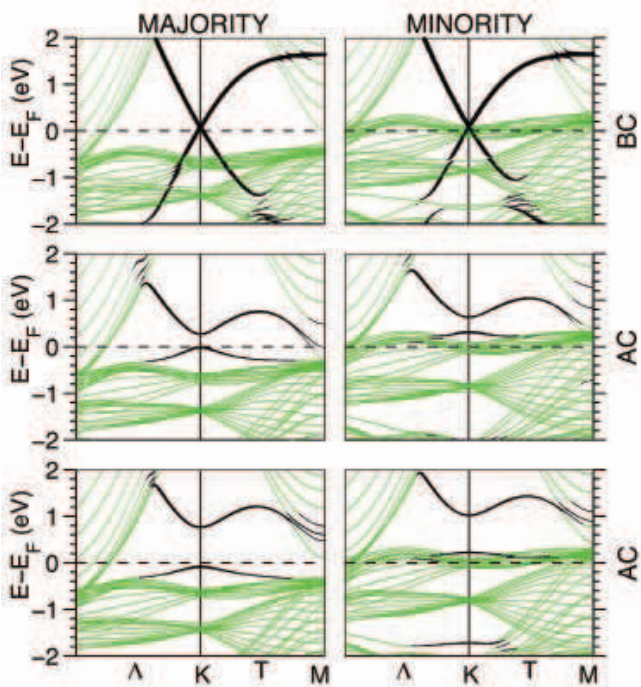}
\caption{ (Color online) The results of PWP (top and middle rows) and
TB-MTO-ASA (bottom row) calculations of majority (left panels) and
minority (right panels) spin band structures (green) of single graphene
layers absorbed on both sides of a 13 layer (111) Ni slab for a BC
configuration with $d_0=3.3$ \AA, (top) and an AC configuration with
$d_0=2.0$ \AA\ (middle and bottom). The bands are replotted and superimposed
in black using the carbon $p_z$ character as a weighting factor. The Fermi
energy is indicated by the horizontal dashed line.}
\label{fig:Gr_TMR_bands}
\end{figure}

The next step is to put a monolayer of graphite (graphene) on top of
the Ni(111) substrate at a distance $d_0$ from the metal surface. From
our studies of the energetics of graphene on TM(111), we found
\cite{Karpan:prl07,Giovannetti:prl08} that the lowest energy
configuration (with $3m$ symmetry) for TM=Ni or Co corresponds to an
``AC'' configuration in which one carbon atom is positioned on top of a
surface TM atom (an ``A'' site) while the second carbon atom is
situated above a third layer TM atom (a ``C'' site), where A and C
refer to the ABC stacking of fcc close-packed planes, see
Fig.~\ref{fig:Gr_TMR_fig3}. This is in agreement with another recent
first-principles calculations \cite{Bertoni:prb05} as well as with
experiments \cite{Gamo:ss97,Oshima:jpcm97} for graphene on the Ni(111)
surface. The electronic structure of a single graphene sheet will
depend on $d_0$ and the details of such graphene-metallic substrate
contacts can be expected to play an important role in current-in-plane
(CIP) devices. \cite{Hill:ieeem06,Tombros:nat07} For the less strongly
bound BC configuration of Gr on Ni, the equilibrium separation is
rather large, $d_0\sim 3.3$ \AA\, and the characteristic band structure
of an isolated graphene sheet is clearly recognizable; see
Fig.~\ref{fig:Gr_TMR_bands}. For the lowest energy AC configuration,
the interaction between the graphene sheet and Ni surface is much
stronger, a gap is opened in the graphene derived $p_z$ bands and at
the Fermi energy there are no graphene states at the K-point in
reciprocal space for the minority spin channel. This may complicate
efficient spin injection into graphene in lateral, CIP devices.
\cite{Hill:ieeem06}

The band structure calculated with the TB-MTO-AS approximation for the
AC configuration is shown in the bottom panel of
Fig.~\ref{fig:Gr_TMR_bands} and is seen to describe graphene on Ni(111)
qualitatively quite well. However, the splitting of the graphene bands,
which arises because the two carbons atoms are no longer equivalent
when one is above a top layer A site Ni atom and the other is above a
third layer C site Ni atom, is somewhat larger than that resulting from
the PWP calculation.

\subsection{${\rm Ni|Gr_n|Ni(111)}$ junction}

The transmission of electrons through a TM$|$Gr$_n|$TM junction will
obviously depend on the geometry of the metal-graphite contacts. Rather
than carrying out a total energy minimization explicitly for every
different value of $n$, we assume that the weak interaction between
graphene sheets will not influence the stronger TM$|$Gr interaction and
construct the junction using the ``AC'' configuration and the
equilibrium separation $d_0=2.03$ \AA\ for each interface, as shown in
Fig.~\ref{fig:Gr_TMR_fig3}. The interstitial space at the TM$|$Gr interfaces is
filled with empty spheres using a procedure analogous to that described
for bulk graphite. \cite{note3}

Because the two carbon atoms c$_1$ and c$_2$ in the graphene unit cell
are equivalent, either of them can be positioned above a surface Ni
atom on an A site with the other on the C site in an ``AC''
configuration, without changing the total energy. Since this can be
done for each TM$|$Gr interface separately, four different
configurations of the TM$|$Gr$|$TM junction can be constructed by
rotating one or both electrodes through $180^{\circ}$ about a vertical
axis through the second layer B sites which interchanges electrode A
and C sites in Fig.~\ref{fig:Gr_TMR_fig3}. We label these four
different configurations c$_1$c$_1$, c$_1$c$_2$, c$_2$c$_1$ and
c$_2$c$_2$ in terms of the carbon atoms which are bonded to A site TM
atoms. For more than one graphene sheet, the second sheet breaks the
symmetry between the c$_1$ and c$_2$ atoms. While we have not checked
this explicitly, we expect the corresponding energy difference to be
small and neglect it.

\begin{figure}[btp]
\includegraphics[scale=0.65]{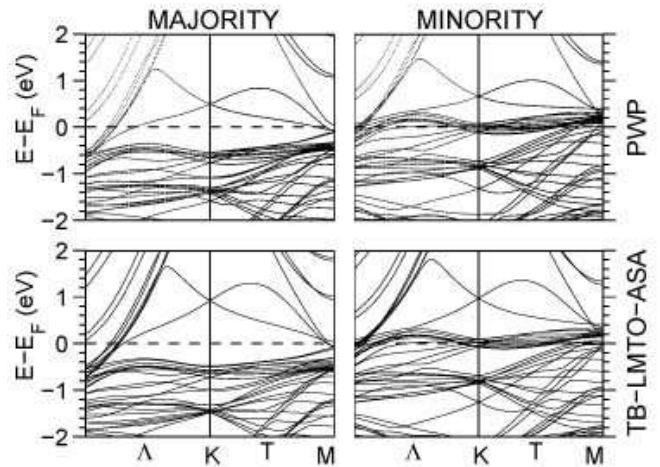}
\caption{Energy band structures of an ideal Ni$_6|$Gr (111) multilayer
with 6 layers of fcc Ni sandwiching a single graphene sheet in a
c$_1$c$_2$ configuration, for majority (left panels) and minority
(right panels) spin channels. Plane wave pseudopotential calculations
are shown on top (dotted lines), the TB-LMTO-ASA results on the
bottom (solid lines).}
\label{fig:Gr_TMR_fig4}
\end{figure}

In Figure~\ref{fig:Gr_TMR_bands} we saw that the graphene $\pi$ states
interacted strongly with the nickel surface in the minimum energy
``AC'' configuration. The interaction with the metal substrate made the
c$_1$ and c$_2$ carbon atoms inequivalent and led to the opening of an
energy gap in the graphene $\pi$ bands. Having constructed an interface
geometry, we study the band structure of the Ni$|$graphene$|$Ni (111)
junction as a function of ${\bf k_{\parallel}}$, the two dimensional
Bloch vector, modelling it as a Ni$_3|$graphene$|$Ni$_3$ junction
repeated periodically in the (111) direction (which is equivalent to
a Ni$_6|$Gr$_1$ multilayer). The bands in the top panels of
Fig.~\ref{fig:Gr_TMR_fig4} were calculated using the benchmark
plane-wave pseudopotential (PWP) method, those in the bottom panels
with the TB-LMTO-ASA. We see that the Ni-related bands are described
well by the TB-LMTO-ASA - as might be expected since the ASA is known
to work well for close-packed solids. The second thing we see is that
there is no gap in the graphene $\pi$ bands. This is because the
c$_1$c$_2$ configuration used in the calculation has inversion symmetry
and the equivalence between the two carbon atoms is restored; see
Fig.~\ref{fig:Gr_TMR_fig3}(a). The third point to be made is that the
charge transfer from graphene (work function: 4.5 eV) to Ni (work
function: 5.5 eV) and strong chemisorption leads to the formation of a
potential step at the interface and a significant shift of the graphene
$\pi$ bands with respect to the Fermi level \cite{Giovannetti:prl08}
which is pinned at that of bulk Ni. We find similar results for
Co$|$Gr$|$Co(111) and Co$|$Gr$|$Co(0001) junctions. There is a
difference between the position of the graphene $\pi$-derived bands,
most noticeably at the K point, in the PWP and TB-LMTO-ASA band
structures shown in Fig.~\ref{fig:Gr_TMR_fig4} for both spin channels.
It appears that the interface dipole is not accurately described by the
ASA. From the point of view of describing transmission of electrons
through this junction, the electronic band structure is the most
important measure of the quality of our basis, description of the
potential, etc., so this discrepancy will most certainly have
quantitative implications. Fortunately, our most important conclusions
will be qualitative and will not depend on this aspect of the
electronic structure.

\section{Electron transport through a $\rm FM|Gr_n|FM$ junction}
\label{sec:Gr_TMR_transport}

Using the geometries and potentials described above, we proceed to
study the spin-dependent transmission through ideal Ni$|$Gr$|$Ni
junctions in the CPP geometry as a function of the thickness of the
graphite spacer layer. We then discuss how interface roughness, alloy
disorder and the lattice mismatch between graphite and the substrate
affect the spin-filtering properties of the junctions using large
lateral supercells to model the various types of disorder. Because very
similar results are found for all the TMs shown in
Fig.~\ref{fig:Gr_TMR_fig5}, we focus on fcc Ni as a substrate because
it has the smallest lattice mismatch with graphite and graphene has
been successfully grown on Ni using chemical vapour deposition.
\cite{Dedkov:apl08,Oshima:jpcm97,Gamo:ss97}

\subsection{Specular interface}

\begin{figure}[!b]
\includegraphics[scale=0.64]{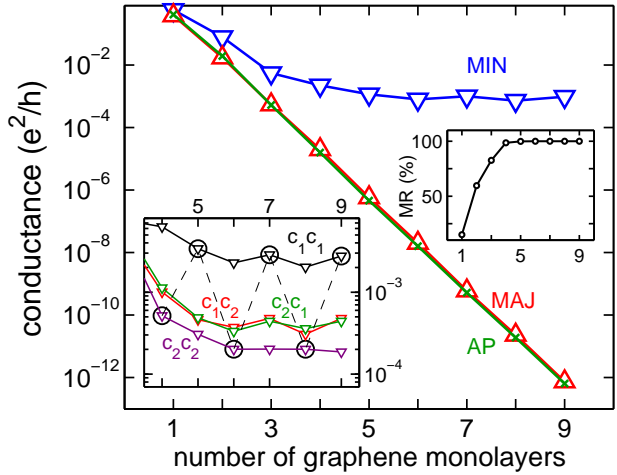}
\caption{ Conductances $G_P^{\rm min}$ ($\triangledown$), $G_P^{\rm
maj}$ ($\vartriangle$), and $G_{AP}^{\protect\sigma}$
(${\mathbf{\times}}$) averaged over the four configurations c$_1$c$_1$,
c$_1$c$_2$, c$_2$c$_1$ and c$_2$c$_2$ of a Ni$|$Gr$_n|$Ni junction as a
function of the number of graphene monolayers $n$ for ideal junctions.
Right inset: magnetoresistance MR as a function of $n$. Left inset:
minority parallel conductance $G_P^{\rm min}$ ($\triangledown$) given
for four different configurations. The points which are circled and
connected with a dashed line are the values which were shown in
Ref.~\onlinecite{Karpan:prl07}.}
\label{fig:Gr_TMR_fig6}
\end{figure}

The spin-dependent transmission through Ni$|$Gr$_n|$Ni (111) junctions
is shown in Fig.~\ref{fig:Gr_TMR_fig6} for parallel (P) and antiparallel (AP)
orientations of the magnetization in the nickel leads, in the form of
the conductances $G_P^{\sigma}$ and $G_{\rm AP}^{\sigma}$ with
$\sigma=$ min, maj. All the conductance values are averaged over the
four interface configurations of the Ni$|$Gr$_n|$Ni junction which are
consistent with AC configurations of the Ni$|$Gr (111) interface.
$G_{\rm P}^{\rm maj}$ and $G_{\rm AP}^{\sigma}$ are strongly
attenuated, while $G_{\rm P}^{\rm min}$ saturates to an $n$-independent
value. The magnetoresistance (MR) defined as
\begin{equation}
{\rm MR} = \frac{R_{\rm AP} - R_{\rm P}}{R_{\rm AP}} \times 100\%
\equiv
\frac{G_{\rm P} - G_{\rm AP}}{G_{\rm P}} \times 100\%,
\label{Eq1}
\end{equation}
rapidly approaches its maximum possible value of 100\%, as shown in the
right inset in Fig.~\ref{fig:Gr_TMR_fig6}. This \emph{pessimistic}
definition of MR is more convenient here because $G_{\rm AP}$ vanishes
for large $n$. It is usually the optimistic version, that approaches
$10^{12}$~\% in our calculations but does not saturate, that is quoted.
\cite{Yuasa:natm04,Parkin:natm04,Butler:prb01,Mathon:prb01} The left
inset in Fig.~\ref{fig:Gr_TMR_fig6} shows how the conductance depends
on the particular configuration of the junction. The minority spin
conductance in the parallel configuration, which dominates the
magnetoresistance behaviour, is highest for the c$_1$c$_1$
configuration with an asymptotic value of $G_{\rm P}^{\rm min} \sim
10^{-2}\, G_0$. This is approximately an order of magnitude larger than
$G_{\rm P}^{\rm min}$ for the c$_2$c$_1$, c$_1$c$_2$ and c$_2$c$_2$
configurations. The c$_1$c$_2$ and c$_2$c$_1$ configurations are
equivalent so the corresponding values of $G_{\rm P}^{\rm min}$ should
be identical. The small differences between these two configurations
which can be seen in the figure are
an indication of the overall accuracy of the numerical calculation. The
points which are circled and connected with a dashed line are the
oscillating values which were shown in Ref.~\onlinecite{Karpan:prl07}.

\begin{figure}[!b]
\includegraphics[scale=0.35]{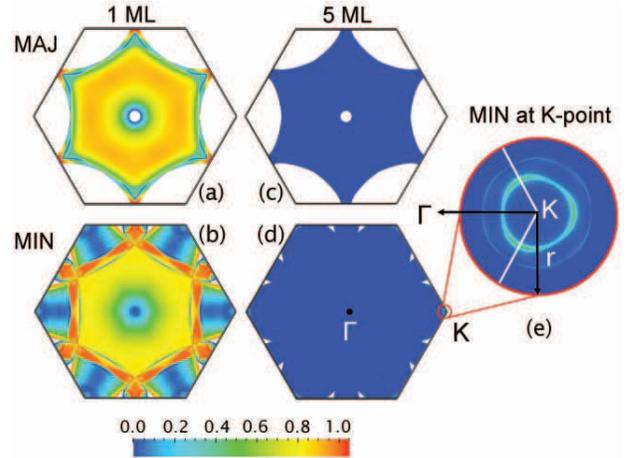}
\caption{ Transmission as a function of the transverse crystal momentum
${\bf k}_{\parallel}$ in the two dimensional interface BZ for a
c$_1$c$_2$ configuration of an ideal Ni$|$Gr$_n|$Ni (111) junction in a
parallel state. (a) and (b) are for a single graphene sheet, $n=1$; (c)
and (d) are for $n=5$; (e) shows the minority spin transmission in a
small circle of radius $r=0.057$ ($2\pi/a_{\rm Gr}$) around the K point
for 5 ML of graphite on an enlarged scale.} \label{fig:Gr_TMR_fig7}
\end{figure}

To demonstrate that spin-filtering occurs due to high transmission of
minority spin electrons around the K point, we plot the majority- and
minority-spin transmission for the P configuration as a function of
${\bf k}_{\parallel}$ for two graphite films of different thickness in
Figure~\ref{fig:Gr_TMR_fig7}. A single sheet of graphene (a monolayer
of graphite) is essentially transparent with a conductance of order
$G_0$ in both spin channels. In the minority spin channel, the
transmission is very low or vanishes close to $\bar{\Gamma}$ and close
to K along the high symmetry $\Gamma$-K line, in spite of there being
one or more sheets of Fermi surface in these regions of reciprocal
space. This is a clear indication of the importance of matrix element
effects: selection rules resulting from the incompatibility of wave
functions on either side of the interface. \cite{Xia:prb06}

The majority transmission must be zero around $\bar{\Gamma}$ and around
the K point because there are no states there in the Ni leads. For
thicker graphite, the only contribution to the majority-spin
conductance comes from tunneling through graphite in regions of the
2D-BZ where there are Ni states and the gap between graphite bonding
and antibonding $\pi$ states is small. This occurs close to the M
point;\cite{Charlier:prb91a} see Fig.~\ref{fig:Gr_TMR_fig2}. Because
the gap decreases going from M to K, the transmission increases in this
direction. At the edge of the Fermi surface projection, the velocity of
the Bloch electrons in the leads is zero so that the maximum
transmission occurs just on the M side of these edges.

The total minority transmission consist of two contributions. On the
one hand there is a tunneling contribution from throughout the 2D-BZ
which, depending on the particular ${\bf k}_{\parallel}$ point, is
determined by the gap in graphite as well as by the compatibility of
the symmetries, at that point, of the wave functions in Ni and in
graphite. On the other hand there is a large transmission from the
neighbourhood of the K point coming from the Bloch states there in
graphite. Once these have coupled to available states in Ni, this
contribution does not change much as more layers of graphite are added.
Perfect spin-filtering (100\% magnetoresistance) occurs when the
tunneling contributions are essentially quenched compared to the
minority spin K point contribution. For four MLs of graphite the
polarization is within a percent of 100\% and for five MLs it is for
all intents and purposes complete. The only discernible transmission in
Figs.~\ref{fig:Gr_TMR_fig7}(c-e) is found close to the K point.
Magnification of this region in Fig.~\ref{fig:Gr_TMR_fig7}(e) shows a
certain amount of structure in the transmission. This can be explained
in terms of the multiple sheets of Ni minority spin Fermi surface in
the vicinity of K (Fig.~\ref{fig:Gr_TMR_fig5}) and the small but finite
dispersion of the graphite bands perpendicular to the basal plane.
\cite{Charlier:prb91a} The transmission is seen to have the threefold
symmetry of the junction.

The spin-filtering does not depend on details of how graphite is bonded
to the ferromagnetic leads as long as the translational symmetry
parallel to the metal-graphite interfaces is preserved. We have
verified this by performing explicit calculations (results not shown
here) for junctions in the ``AB'' and ``BC'' configurations with
different metal-graphite separations $d$.

\subsection{  Ni$|$Cu$_m|$Gr$_n|$Cu$_m|$Ni (111)  }
In Section~\ref{sec:Gr_TMR_geometry}, we saw that the electronic
structure of a sheet of graphene depends strongly on its separation
from the underlying TM substrate. For Co and Ni, equilibrium
separations of the order of 2.0 \AA\ were calculated for the lowest
energy AC configuration (see Table~\ref{tab:Gr_TMR_tableone}), the interaction
was strong and the characteristic linear dispersion of the graphene
electronic structure was destroyed, Fig.~\ref{fig:Gr_TMR_bands}. For a separation
of 3.3 \AA, the small residual interaction does not destroy the linear
dispersion. Unlike Co and Ni, Cu interacts only weakly with graphene,
there is only a small energy difference between the ``asymmetric'' AC
configuration with $d_0=3.3$~\AA\ and the slightly more weakly bound
``symmetric'' BC configuration with $d_0=3.4$~\AA, and bonding to Cu
preserves the characteristic graphene electronic structure, opening up
only a very small gap of about 10 meV at the Dirac point.
\cite{Giovannetti:prb07}

\begin{figure}[!b]
\includegraphics[scale=0.37]{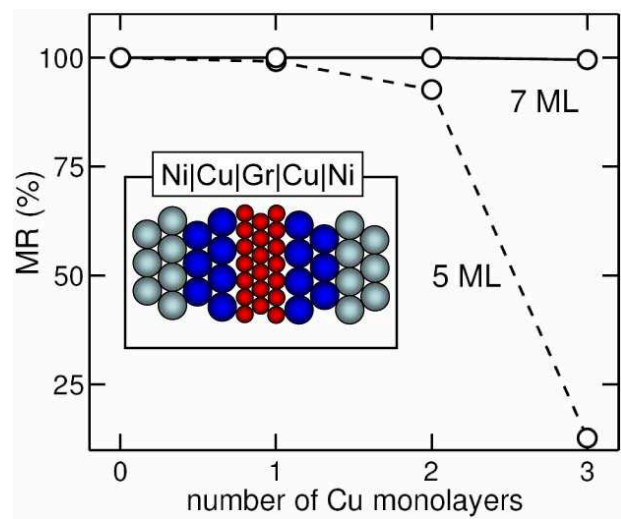}
\caption{Magnetoresistance as a function of the number of Cu monolayers
on both left and right Ni leads in case of 5 ML (dashed line) and 7 ML
(solid line) of graphene.} \label{fig:Gr_TMR_fig8}
\end{figure}

Should it be desirable to avoid forming a strong bond between graphite
and the TM electrode, then it should be a simple matter of depositing
one or a few layers of Cu on e.g. Ni. Such a thin layer of Cu will
adopt the in-plane lattice constant of Ni and graphite will bind to it
weakly so that the electronic structure of the first layer of graphite
will be only weakly perturbed. Because Cu oxidizes less readily than Ni
or Co, it may be used as a protective layer. Cu has no states at or
around the K point for either spin channel (Fig.~\ref{fig:Gr_TMR_fig5})
so it will simply attenuate the conductance of the minority spin
channel at the K point. This is demonstrated in
Fig~\ref{fig:Gr_TMR_fig8} where the magnetoresistance of a
Ni$|$Cu$_m|$Gr$_n|$Cu$_m|$Ni junction is shown as a function of the
number $m$ of layers of Cu when there are 5 MLs and 7 MLs of graphite.
As the thickness of Cu is increased reducing the transmission of the
minority-spin K point channel, the MR decreases. The reduction of the
MR can be compensated by increasing the thickness of graphite. These
conclusions are consistent with the qualitative conclusions drawn above
in connection with Fig~\ref{fig:Gr_TMR_fig5}(a).

Although the linear dispersion of the graphene bands is essentially
unchanged by adsorption on Cu, application of an in-plane bias will
destroy the translational symmetry parallel to the interface upon which
our considerations have been based. The finite lateral size of a
Ni$|$Cu electrode will also break the translational symmetry in a CIP
measuring configuration and edge effects may destroy the spin-injection
properties.

\subsection{Effect of disorder}
\subsubsection{Lattice Mismatch}
So far, we have assumed TM and graphite lattices which are commensurate
in-plane. In practice there is a lattice mismatch with graphite of
1.3\% for Ni, 1.9\% for Co and 3.9\% for Cu which immediately poses the
question of how this will affect the perfect spin-filtering. While
lattice mismatch between lattices with lattice constants $a_1$ and
$a_2$ can in principle be treated by using $n_1$ units of lattice 1 and
$n_2$ units of lattice 2 with $n_1 a_1=n_2 a_2$, in practice we cannot
perform calculations for systems with $n$ much larger than 20 which
limits us to treating a large lattice mismatch of 5\%. To put an upper
limit on the effect of a $1.3-1.9\%$ lattice mismatch, we performed
calculations for a Ni$|$Gr$_5|$Ni junction matching $19 \times 19$ unit
cells of Ni in-plane to $20 \times 20$ unit cells of graphite. The
effect of this 5\% lattice mismatch was to reduce the (pessimistic)
magnetoresistance from 100\% to 90\% (or $\sim 900$\% in the optimistic
definition). We conclude that the actual Ni$|$Gr mismatch of 1.3\%
should not be a serious limiting factor in practice.

\subsubsection{Interface Roughness}
Incommensurability is not the only factor that might reduce the
magnetoresistance. Preparing atomically perfect interfaces is not
possible and raises the question of how sensitive the perfect
spin-filtering will be to interface roughness or disorder. Our studies
of spin injection in Ref.~\onlinecite{Zwierzycki:prb03} and TMR in
Ref.~\onlinecite{Xu:prb06} suggest they may be very important and can
even dominate the spin transport properties.

The simplest way to prepare a CPP Ni$|$Gr$|$Ni junction would
presumably be to begin with a (111) oriented Ni or Co crystal
characterized on an atomic scale by STM or AFM, grow the required
number of layers of graphene by e.g. chemical vapour deposition
\cite{Dedkov:apl08,Oshima:jpcm97,Gamo:ss97} and after characterization
of the graphene layers to then deposit the second Ni electrode. To
prepare a CIP junction, we envisage a procedure in which thin graphite
layers are prepared by micromechanical cleavage of bulk graphite onto a
SiO$_2$ covered Si wafer \cite{Geim:natm07} into which TM (Ni or Co)
electrodes have been embedded. We assume that the (111) electrodes can
be prepared in ultrahigh vacuum and characterized on an atomic scale
and that the surfaces are flat and defect free. Layers of graphene are
peeled away until the desired value of $n$ is reached.

\begin{figure}[btp]
\includegraphics[scale=0.37]{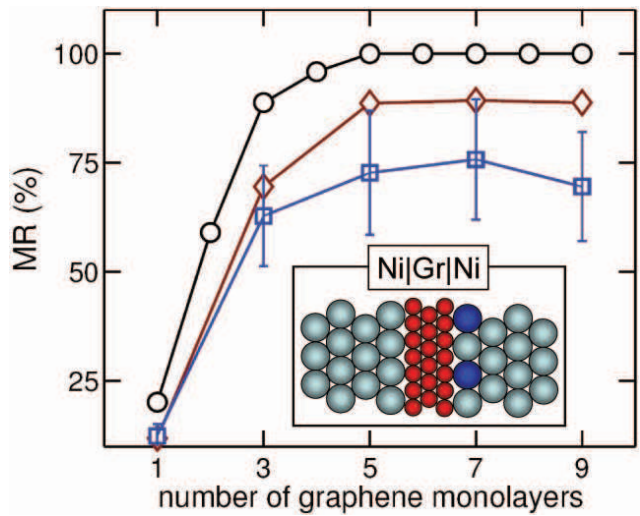}
\caption{ (Color online) Magnetoresistance as a function of $n$ for:
ideal junctions (circles); Ni$|$Gr$_n|$Cu$_{50}$Ni$_{50}|$Ni junctions where
the surface layer is a disordered alloy (diamonds); Ni$|$Gr$_n|$Ni
junctions where the top layer of one of the electrodes is rough with
only half of the top layer sites occupied (squares). For the rough
surface layer, the error bars
indicate the spread of MR obtained for different configurations. Inset:
schematic representation of Ni$|$Gr$_n|$Ni junction with alloy disorder
(roughness) at the right Ni$|$Gr interface. Ni atoms are given by large
gray spheres while Cu (missing) atoms in the case of alloy disorder
(roughness) are given by large dark (blue) spheres. Positions of carbon
atoms are represented by small dark (red) spheres.}
\label{fig:Gr_TMR_fig9}
\end{figure}

Assuming it will be possible to realize one essentially perfect
interface, we have studied the effect of roughness at the second
interface, assuming it is prepared by evaporation or some similar
method. The graphite is assumed to be atomically perfect and all of the
roughness occurs in the metal interface layer. We model this roughness
as in Ref.~\onlinecite{Xu:prb06} by removing a certain percentage of
the top layer atoms. The atomic sphere potentials are calculated using
the layer version \cite{Turek:97} of the coherent potential
approximation (CPA). \cite{Soven:pr67} The CPA AS potentials are then
distributed at random with the appropriate concentration in $5 \times
5$ lateral supercells and the transmission is calculated in a CPP
geometry for a number of such randomly generated configurations. The
effect on the magnetoresistance of removing half a monolayer of Ni is
shown in Fig.~\ref{fig:Gr_TMR_fig9} as a function of the number of
graphite layers. 50\% roughness at one interface is seen to reduce the
100\% magnetoresistance to about 70\% ($\sim 230\%$ optimistic).

\subsubsection{Interface Disorder}
The last type of disorder we consider is a layer of interface alloy. We
imagine that depositing a layer of Cu on Ni to prevent graphite bonding
to the Ni has led to a layer of Ni and Cu mixing. In a worst case
scenario, we assume all of the disorder is in the surface layer and
assume this to be a Ni$_{50}$Cu$_{50}$ random alloy. The potentials are
once again calculated self-consistently using the layer CPA and the
transmission calculated as for roughness. The effect on a monolayer of
CuNi alloy is to reduce the MR to 90\% (900\% in the optimistic
definition) for a thick graphite film, as shown in
Fig.~\ref{fig:Gr_TMR_fig9}. These results indicate that the momentum
transfer induced by the scattering due to imperfections is insufficient
to bridge the large gap about the K point in the majority spin FS
projections.

Ideally, we should avoid interface roughness and disorder altogether.
Since metal surfaces can be prepared with very little disorder, what is
required is to be able to perform micromechanical cleavage on a metal
surface rather than on SiO$_2$. If this were possible, two essentially
perfect TM$|$Gr interfaces could perhaps be joined using a method
analogous to vacuum bonding. \cite{Monsma:sc98} Alternatively, since
graphite has a large $c$-axis resistivity \cite{Matsubara:prb90} it may
only be necessary to prepare one near-perfect Ni$|$graphite interface.
If the graphite layer is sufficiently thick, then it should be possible
to achieve 100\% spin accumulation in a high resistivity material
making it suitable for injecting spins into semiconductors.
\cite{Schmidt:prb00} Because carbon is so light, spin-flip scattering
arising from spin-orbit interaction should be negligible.

\section{Discussion and Conclusions}
\label{sec:Gr_TMR_conclusions}

Motivated by the recent progress in preparing and manipulating
discrete, essentially atomically perfect graphene layers, we have used
parameter-free, materials-specific electronic structure calculations to
explore the spin transport properties of a novel TM$|$graphite system.
Perfect spin-filtering is predicted for ideal TM$|$Gr$_n|$TM junctions
with TM = Co or Ni in both fcc and hcp crystal structures. The spin
filtering stems from a combination of almost perfect matching of Gr and
TM lattice constants and unique features of their electronic band
structures. Graphite films have occupied states at the Fermi level only
around the K-point in the first (interface) BZ. Close-packed fcc and
hcp Ni and Co have only minority spin states in the vicinity of the
same K point, at the Fermi energy. For a modest number of layers of
graphite, transport from one TM electrode to the other can only occur
via the graphite states close to the K point and perfect spin filtering
occurs if the in-plane translational symmetry is preserved. For
majority spins, the graphite film acts as a tunnel barrier while it is
conducting for minority spin electrons, albeit with a small
conductance.
\begin{figure}[!t]
\includegraphics[scale=0.35]{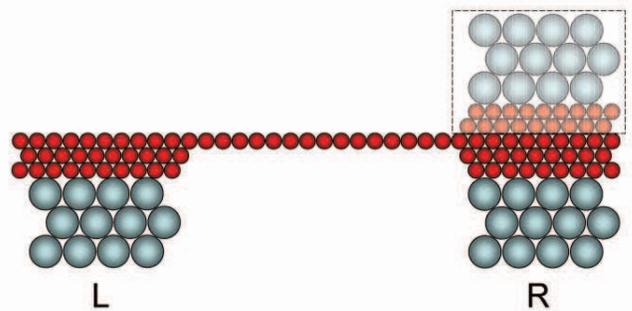}
\caption{Schematic figure of a TM$|$Gr$|$TM CIP junction in which the
electric field is forced to be essentially perpendicular to the
TM$|$graphene interface. The dashed shaded box indicates an alternative
configuration with the right-hand electrode on top of the graphene.}
\label{fig:Gr_fig11}
\end{figure}

Compared to a conventional magnetic tunnel junction, a TM$|$Gr$_n|$TM
CPP junction has several important advantages. Firstly, the lateral
lattice mismatch is three times smaller than the 3.8\% found for the
now very popular Fe$|$MgO$|$Fe(001) MTJs.\cite{Yuasa:jpd07} This will
reduce the number of defects caused by strain that otherwise limits the
thickness of the tunnel barrier and degrades the efficiency of spin
injection. Secondly, the spin polarization approaches 100\% for an
ideal junction with $n>3$ graphene layers, and is only reduced to
70-90\% for junctions with large interface roughness or disorder.
Thirdly, the spin-filtering effect should not be very sensitive to
temperature. From Fig.~\ref{fig:Gr_TMR_fig7}(a) and the corresponding
figures for other thicknesses of graphite, we see that the largest
contribution to the majority spin conduction comes from tunneling at
the M point where bulk Co and Ni have propagating states at the Fermi
level and the distance in energy to states in graphite with the same
${\bf k_{\parallel}}$ vector is a minimum.  From
Fig.~\ref{fig:Gr_TMR_fig2} we see that the energy gap is almost 1 eV
between the Fermi level and the closest graphite band at this point. To
bridge the horizontal gap between states close to the K-point in
graphite and the closest states in Co or Ni requires an in-plane
momentum transfer of order $\Delta k \sim \pi/a$. The corresponding
energy would be (comparable to) that of an optical phonon which is
large because of the stiffness of a graphene sheet.

To achieve perfect spin-injection into a single sheet of graphene is
more troublesome. \cite{Hill:ieeem06,Tombros:nat07} The electronic
structure calculations presented here show that the carbon $\pi$
orbitals hybridize strongly with Ni (and also Co) surfaces leading to
the destruction of graphene's characteristic electronic structure. We
have already suggested that dusting Ni (or Co) with Cu will lead to
near-complete restoration of the graphene electronic structure because
of the weak interaction between graphene and Cu. Moreover Cu might also
prevent rapid oxidation of the Ni(Co) (111) surfaces, which could be
important for making practical devices. However, application of a bias
would lead to a breaking of the translational symmetry responsible for
the perfect spin filtering. The finite size of electrodes might also
present a problem in practice especially if the potential drop occurs
at the edges. The problem can be simply solved by forcing the electric
field to be perpendicular to the TM$|$graphite interface as sketched in
Fig.~\ref{fig:Gr_fig11} where the right electrode could equally well be
placed on top of the graphite.

In conclusion, we propose a new class of lattice-matched junctions,
TM$|$Gr$_{n}|$TM, that exhibit exceptionally high magnetoresistance
effect which is robust with respect to interface disorder, roughness,
and finite temperatures making them highly attractive for possible
applications in spintronic devices.

\acknowledgments

This work is supported by ``NanoNed'', a nanotechnology programme of
the Dutch Ministry of Economic Affairs. It is part of the research programs
of ``Chemische Wetenschappen'' (CW) and ``Stichting voor Fundamenteel
Onderzoek der Materie'' (FOM) and the use of supercomputer facilities
was sponsored by the ``Stichting Nationale Computer
Faciliteiten'' (NCF), all financially supported by the ``Nederlandse
Organisatie voor Wetenschappelijk Onderzoek'' (NWO).
MZ wishes to acknowledge support from EU grant CARDEQ under contract
IST-021285-2.



\begin{thebibliography}{60}
\expandafter\ifx\csname natexlab\endcsname\relax\def\natexlab#1{#1}\fi
\expandafter\ifx\csname bibnamefont\endcsname\relax
  \def\bibnamefont#1{#1}\fi
\expandafter\ifx\csname bibfnamefont\endcsname\relax
  \def\bibfnamefont#1{#1}\fi
\expandafter\ifx\csname citenamefont\endcsname\relax
  \def\citenamefont#1{#1}\fi
\expandafter\ifx\csname url\endcsname\relax
  \def\url#1{\texttt{#1}}\fi
\expandafter\ifx\csname urlprefix\endcsname\relax\def\urlprefix{URL }\fi
\providecommand{\bibinfo}[2]{#2}
\providecommand{\eprint}[2][]{\url{#2}}

\bibitem[{\citenamefont{Karpan et~al.}(2007)\citenamefont{Karpan, Giovannetti,
  Khomyakov, Talanana, Starikov, Zwierzycki, van~den Brink, Brocks, and
  Kelly}}]{Karpan:prl07}
\bibinfo{author}{\bibfnamefont{V.~M.} \bibnamefont{Karpan}},
  \bibinfo{author}{\bibfnamefont{G.}~\bibnamefont{Giovannetti}},
  \bibinfo{author}{\bibfnamefont{P.~A.} \bibnamefont{Khomyakov}},
  \bibinfo{author}{\bibfnamefont{M.}~\bibnamefont{Talanana}},
  \bibinfo{author}{\bibfnamefont{A.~A.} \bibnamefont{Starikov}},
  \bibinfo{author}{\bibfnamefont{M.}~\bibnamefont{Zwierzycki}},
  \bibinfo{author}{\bibfnamefont{J.}~\bibnamefont{van~den Brink}},
  \bibinfo{author}{\bibfnamefont{G.}~\bibnamefont{Brocks}}, \bibnamefont{and}
  \bibinfo{author}{\bibfnamefont{P.~J.} \bibnamefont{Kelly}},
  \bibinfo{journal}{Phys. Rev. Lett.} \textbf{\bibinfo{volume}{99}},
  \bibinfo{pages}{176602} (\bibinfo{year}{2007}).

\bibitem[{\citenamefont{\v{Z}uti\'c et~al.}(2004)\citenamefont{\v{Z}uti\'c,
  Fabian, and Sarma}}]{Zutic:rmp04}
\bibinfo{author}{\bibfnamefont{I.}~\bibnamefont{\v{Z}uti\'c}},
  \bibinfo{author}{\bibfnamefont{J.}~\bibnamefont{Fabian}}, \bibnamefont{and}
  \bibinfo{author}{\bibfnamefont{S.~D.} \bibnamefont{Sarma}},
  \bibinfo{journal}{Rev. Mod. Phys.} \textbf{\bibinfo{volume}{76}},
  \bibinfo{pages}{323} (\bibinfo{year}{2004}).

\bibitem[{\citenamefont{Neto et~al.}(2008)\citenamefont{Neto, Guinea, Peres,
  Novoselov, and Geim}}]{Neto:rmp08}
\bibinfo{author}{\bibfnamefont{A.~H.~C.} \bibnamefont{Neto}},
  \bibinfo{author}{\bibfnamefont{F.}~\bibnamefont{Guinea}},
  \bibinfo{author}{\bibfnamefont{N.~M.~R.} \bibnamefont{Peres}},
  \bibinfo{author}{\bibfnamefont{K.~S.} \bibnamefont{Novoselov}},
  \bibnamefont{and} \bibinfo{author}{\bibfnamefont{A.~K.} \bibnamefont{Geim}},
  \bibinfo{journal}{Rev. Mod. Phys.} \textbf{\bibinfo{volume}{--}},
  (\bibinfo{year}{2008}), \bibinfo{note}{cond-mat/0709.1163}.

\bibitem[{\citenamefont{Baibich et~al.}(1988)\citenamefont{Baibich, Broto,
  Fert, {Nguyen Van Dau}, Petroff, Etienne, Creuzet, Friederich, and
  Chazelas}}]{Baibich:prl88}
\bibinfo{author}{\bibfnamefont{M.~N.} \bibnamefont{Baibich}},
  \bibinfo{author}{\bibfnamefont{J.~M.} \bibnamefont{Broto}},
  \bibinfo{author}{\bibfnamefont{A.}~\bibnamefont{Fert}},
  \bibinfo{author}{\bibfnamefont{F.}~\bibnamefont{{Nguyen Van Dau}}},
  \bibinfo{author}{\bibfnamefont{F.}~\bibnamefont{Petroff}},
  \bibinfo{author}{\bibfnamefont{P.}~\bibnamefont{Etienne}},
  \bibinfo{author}{\bibfnamefont{G.}~\bibnamefont{Creuzet}},
  \bibinfo{author}{\bibfnamefont{A.}~\bibnamefont{Friederich}},
  \bibnamefont{and} \bibinfo{author}{\bibfnamefont{J.}~\bibnamefont{Chazelas}},
  \bibinfo{journal}{Phys. Rev. Lett.} \textbf{\bibinfo{volume}{61}},
  \bibinfo{pages}{2472} (\bibinfo{year}{1988}).

\bibitem[{\citenamefont{Binasch et~al.}(1989)\citenamefont{Binasch,
  Gr{\"{u}}nberg, Saurenbach, and Zinn}}]{Binasch:prb89}
\bibinfo{author}{\bibfnamefont{G.}~\bibnamefont{Binasch}},
  \bibinfo{author}{\bibfnamefont{P.}~\bibnamefont{Gr{\"{u}}nberg}},
  \bibinfo{author}{\bibfnamefont{F.}~\bibnamefont{Saurenbach}},
  \bibnamefont{and} \bibinfo{author}{\bibfnamefont{W.}~\bibnamefont{Zinn}},
  \bibinfo{journal}{Phys. Rev. B} \textbf{\bibinfo{volume}{39}},
  \bibinfo{pages}{4828} (\bibinfo{year}{1989}).

\bibitem[{\citenamefont{Julliere}(1975)}]{Julliere:pla75}
\bibinfo{author}{\bibfnamefont{M.}~\bibnamefont{Julliere}},
  \bibinfo{journal}{Phys. Lett. A} \textbf{\bibinfo{volume}{54}},
  \bibinfo{pages}{225} (\bibinfo{year}{1975}).

\bibitem[{\citenamefont{Moodera et~al.}(1995)\citenamefont{Moodera, Kinder,
  Wong, and Meservey}}]{Moodera:prl95}
\bibinfo{author}{\bibfnamefont{J.~S.} \bibnamefont{Moodera}},
  \bibinfo{author}{\bibfnamefont{L.~R.} \bibnamefont{Kinder}},
  \bibinfo{author}{\bibfnamefont{T.~M.} \bibnamefont{Wong}}, \bibnamefont{and}
  \bibinfo{author}{\bibfnamefont{R.}~\bibnamefont{Meservey}},
  \bibinfo{journal}{Phys. Rev. Lett.} \textbf{\bibinfo{volume}{74}},
  \bibinfo{pages}{3273} (\bibinfo{year}{1995}).

\bibitem[{\citenamefont{Miyazaki and Tezuka}(1995)}]{Miyazaki:jmmm95}
\bibinfo{author}{\bibfnamefont{T.}~\bibnamefont{Miyazaki}} \bibnamefont{and}
  \bibinfo{author}{\bibfnamefont{N.}~\bibnamefont{Tezuka}},
  \bibinfo{journal}{J. Magn. \& Magn. Mater.} \textbf{\bibinfo{volume}{139}},
  \bibinfo{pages}{L231} (\bibinfo{year}{1995}).

\bibitem[{\citenamefont{Yuasa et~al.}(2004)\citenamefont{Yuasa, Nagahama,
  Fukushima, Suzuki, and Ando}}]{Yuasa:natm04}
\bibinfo{author}{\bibfnamefont{S.}~\bibnamefont{Yuasa}},
  \bibinfo{author}{\bibfnamefont{T.}~\bibnamefont{Nagahama}},
  \bibinfo{author}{\bibfnamefont{A.}~\bibnamefont{Fukushima}},
  \bibinfo{author}{\bibfnamefont{Y.}~\bibnamefont{Suzuki}}, \bibnamefont{and}
  \bibinfo{author}{\bibfnamefont{K.}~\bibnamefont{Ando}},
  \bibinfo{journal}{Nature Materials} \textbf{\bibinfo{volume}{3}},
  \bibinfo{pages}{868} (\bibinfo{year}{2004}).

\bibitem[{\citenamefont{Parkin et~al.}(2004)\citenamefont{Parkin, Kaiser,
  Panchula, Rice, Hughes, Samant, and Yang}}]{Parkin:natm04}
\bibinfo{author}{\bibfnamefont{S.~S.~P.} \bibnamefont{Parkin}},
  \bibinfo{author}{\bibfnamefont{C.}~\bibnamefont{Kaiser}},
  \bibinfo{author}{\bibfnamefont{A.}~\bibnamefont{Panchula}},
  \bibinfo{author}{\bibfnamefont{P.~M.} \bibnamefont{Rice}},
  \bibinfo{author}{\bibfnamefont{B.}~\bibnamefont{Hughes}},
  \bibinfo{author}{\bibfnamefont{M.}~\bibnamefont{Samant}}, \bibnamefont{and}
  \bibinfo{author}{\bibfnamefont{S.~H.} \bibnamefont{Yang}},
  \bibinfo{journal}{Nature Materials} \textbf{\bibinfo{volume}{3}},
  \bibinfo{pages}{862} (\bibinfo{year}{2004}).

\bibitem[{\citenamefont{Yuasa et~al.}(2005)\citenamefont{Yuasa, Katayama,
  Nagahama, Fukushima, Kubota, Suzuki, and Ando}}]{Yuasa:apl05a}
\bibinfo{author}{\bibfnamefont{S.}~\bibnamefont{Yuasa}},
  \bibinfo{author}{\bibfnamefont{T.}~\bibnamefont{Katayama}},
  \bibinfo{author}{\bibfnamefont{T.}~\bibnamefont{Nagahama}},
  \bibinfo{author}{\bibfnamefont{A.}~\bibnamefont{Fukushima}},
  \bibinfo{author}{\bibfnamefont{H.}~\bibnamefont{Kubota}},
  \bibinfo{author}{\bibfnamefont{Y.}~\bibnamefont{Suzuki}}, \bibnamefont{and}
  \bibinfo{author}{\bibfnamefont{K.}~\bibnamefont{Ando}},
  \bibinfo{journal}{Appl. Phys. Lett.} \textbf{\bibinfo{volume}{87}},
  \bibinfo{pages}{222508} (\bibinfo{year}{2005}).

\bibitem[{\citenamefont{Lee et~al.}(2007)\citenamefont{Lee, Hayakawa, Ikeda,
  Matsukura, and Ohno}}]{Lee:apl07}
\bibinfo{author}{\bibfnamefont{Y.~M.} \bibnamefont{Lee}},
  \bibinfo{author}{\bibfnamefont{J.}~\bibnamefont{Hayakawa}},
  \bibinfo{author}{\bibfnamefont{S.}~\bibnamefont{Ikeda}},
  \bibinfo{author}{\bibfnamefont{F.}~\bibnamefont{Matsukura}},
  \bibnamefont{and} \bibinfo{author}{\bibfnamefont{H.}~\bibnamefont{Ohno}},
  \bibinfo{journal}{Appl. Phys. Lett.} \textbf{\bibinfo{volume}{90}},
  \bibinfo{pages}{212507} (\bibinfo{year}{2007}).

\bibitem[{\citenamefont{Yuasa and Djayaprawira}(2007)}]{Yuasa:jpd07}
\bibinfo{author}{\bibfnamefont{S.}~\bibnamefont{Yuasa}} \bibnamefont{and}
  \bibinfo{author}{\bibfnamefont{D.~D.} \bibnamefont{Djayaprawira}},
  \bibinfo{journal}{J. Phys. D: Appl. Phys.} \textbf{\bibinfo{volume}{40}},
  \bibinfo{pages}{R337} (\bibinfo{year}{2007}).

\bibitem[{\citenamefont{Xu et~al.}(2006)\citenamefont{Xu, Karpan, Xia,
  Zwierzycki, Marushchenko, and Kelly}}]{Xu:prb06}
\bibinfo{author}{\bibfnamefont{P.~X.} \bibnamefont{Xu}},
  \bibinfo{author}{\bibfnamefont{V.~M.} \bibnamefont{Karpan}},
  \bibinfo{author}{\bibfnamefont{K.}~\bibnamefont{Xia}},
  \bibinfo{author}{\bibfnamefont{M.}~\bibnamefont{Zwierzycki}},
  \bibinfo{author}{\bibfnamefont{I.}~\bibnamefont{Marushchenko}},
  \bibnamefont{and} \bibinfo{author}{\bibfnamefont{P.~J.} \bibnamefont{Kelly}},
  \bibinfo{journal}{Phys. Rev. B} \textbf{\bibinfo{volume}{73}},
  \bibinfo{pages}{180402(R)} (\bibinfo{year}{2006}).

\bibitem[{\citenamefont{Zwierzycki et~al.}(2003)\citenamefont{Zwierzycki, Xia,
  Kelly, Bauer, and Turek}}]{Zwierzycki:prb03}
\bibinfo{author}{\bibfnamefont{M.}~\bibnamefont{Zwierzycki}},
  \bibinfo{author}{\bibfnamefont{K.}~\bibnamefont{Xia}},
  \bibinfo{author}{\bibfnamefont{P.~J.} \bibnamefont{Kelly}},
  \bibinfo{author}{\bibfnamefont{G.~E.~W.} \bibnamefont{Bauer}},
  \bibnamefont{and} \bibinfo{author}{\bibfnamefont{I.}~\bibnamefont{Turek}},
  \bibinfo{journal}{Phys. Rev. B} \textbf{\bibinfo{volume}{67}},
  \bibinfo{pages}{092401} (\bibinfo{year}{2003}).

\bibitem[{\citenamefont{Wallace}(1947)}]{Wallace:pr47}
\bibinfo{author}{\bibfnamefont{P.~R.} \bibnamefont{Wallace}},
  \bibinfo{journal}{Phys. Rev.} \textbf{\bibinfo{volume}{71}},
  \bibinfo{pages}{622} (\bibinfo{year}{1947}).

\bibitem[{\citenamefont{Slonczewski and Weiss}(1958)}]{Slonczewski:pr58}
\bibinfo{author}{\bibfnamefont{J.~C.} \bibnamefont{Slonczewski}}
  \bibnamefont{and} \bibinfo{author}{\bibfnamefont{P.~R.} \bibnamefont{Weiss}},
  \bibinfo{journal}{Phys. Rev.} \textbf{\bibinfo{volume}{109}},
  \bibinfo{pages}{272} (\bibinfo{year}{1958}).

\bibitem[{\citenamefont{Lomer}(1955)}]{Lomer:prsla55}
\bibinfo{author}{\bibfnamefont{W.~M.} \bibnamefont{Lomer}},
  \bibinfo{journal}{Proc. R. Soc. London, Ser. A}
  \textbf{\bibinfo{volume}{227}}, \bibinfo{pages}{330} (\bibinfo{year}{1955}).

\bibitem[{\citenamefont{Ando}(2005)}]{Ando:jpsj05}
\bibinfo{author}{\bibfnamefont{T.}~\bibnamefont{Ando}}, \bibinfo{journal}{J.
  Phys. Soc. Jpn.} \textbf{\bibinfo{volume}{74}}, \bibinfo{pages}{777}
  (\bibinfo{year}{2005}).

\bibitem[{\citenamefont{Novoselov
  et~al.}(2005{\natexlab{a}})\citenamefont{Novoselov, Jiang, Schedin, Booth,
  Khotkevich, Morozov, and Geim}}]{Novoselov:pnas05}
\bibinfo{author}{\bibfnamefont{K.~S.} \bibnamefont{Novoselov}},
  \bibinfo{author}{\bibfnamefont{D.}~\bibnamefont{Jiang}},
  \bibinfo{author}{\bibfnamefont{F.}~\bibnamefont{Schedin}},
  \bibinfo{author}{\bibfnamefont{T.~J.} \bibnamefont{Booth}},
  \bibinfo{author}{\bibfnamefont{V.~V.} \bibnamefont{Khotkevich}},
  \bibinfo{author}{\bibfnamefont{S.~V.} \bibnamefont{Morozov}},
  \bibnamefont{and} \bibinfo{author}{\bibfnamefont{A.~K.} \bibnamefont{Geim}},
  \bibinfo{journal}{Proc. Natl. Acad. Sci. U.S.A.}
  \textbf{\bibinfo{volume}{102}}, \bibinfo{pages}{10451}
  (\bibinfo{year}{2005}{\natexlab{a}}).

\bibitem[{\citenamefont{Novoselov et~al.}(2004)\citenamefont{Novoselov, Geim,
  Morozov, Jiang, Zhang, Dubonos, Grigorieva, and Firsov}}]{Novoselov:sc04}
\bibinfo{author}{\bibfnamefont{K.~S.} \bibnamefont{Novoselov}},
  \bibinfo{author}{\bibfnamefont{A.~K.} \bibnamefont{Geim}},
  \bibinfo{author}{\bibfnamefont{S.~V.} \bibnamefont{Morozov}},
  \bibinfo{author}{\bibfnamefont{D.}~\bibnamefont{Jiang}},
  \bibinfo{author}{\bibfnamefont{Y.}~\bibnamefont{Zhang}},
  \bibinfo{author}{\bibfnamefont{S.~V.} \bibnamefont{Dubonos}},
  \bibinfo{author}{\bibfnamefont{I.~V.} \bibnamefont{Grigorieva}},
  \bibnamefont{and} \bibinfo{author}{\bibfnamefont{A.~A.}
  \bibnamefont{Firsov}}, \bibinfo{journal}{Science}
  \textbf{\bibinfo{volume}{306}}, \bibinfo{pages}{666} (\bibinfo{year}{2004}).

\bibitem[{\citenamefont{Novoselov
  et~al.}(2005{\natexlab{b}})\citenamefont{Novoselov, Geim, Morozov, Jiang,
  Katsnelson, Grigorieva, Dubonos, and Firsov}}]{Novoselov:nat05}
\bibinfo{author}{\bibfnamefont{K.~S.} \bibnamefont{Novoselov}},
  \bibinfo{author}{\bibfnamefont{A.~K.} \bibnamefont{Geim}},
  \bibinfo{author}{\bibfnamefont{S.~V.} \bibnamefont{Morozov}},
  \bibinfo{author}{\bibfnamefont{D.}~\bibnamefont{Jiang}},
  \bibinfo{author}{\bibfnamefont{M.~I.} \bibnamefont{Katsnelson}},
  \bibinfo{author}{\bibfnamefont{I.~V.} \bibnamefont{Grigorieva}},
  \bibinfo{author}{\bibfnamefont{S.~V.} \bibnamefont{Dubonos}},
  \bibnamefont{and} \bibinfo{author}{\bibfnamefont{A.~A.}
  \bibnamefont{Firsov}}, \bibinfo{journal}{Nature}
  \textbf{\bibinfo{volume}{438}}, \bibinfo{pages}{197}
  (\bibinfo{year}{2005}{\natexlab{b}}).

\bibitem[{\citenamefont{Zhang et~al.}(2005)\citenamefont{Zhang, Tan, Stormer,
  and Kim}}]{Zhang:nat05}
\bibinfo{author}{\bibfnamefont{Y.~B.} \bibnamefont{Zhang}},
  \bibinfo{author}{\bibfnamefont{Y.~W.} \bibnamefont{Tan}},
  \bibinfo{author}{\bibfnamefont{H.~L.} \bibnamefont{Stormer}},
  \bibnamefont{and} \bibinfo{author}{\bibfnamefont{P.}~\bibnamefont{Kim}},
  \bibinfo{journal}{Nature} \textbf{\bibinfo{volume}{438}},
  \bibinfo{pages}{201} (\bibinfo{year}{2005}).

\bibitem[{\citenamefont{Heersche et~al.}(2007)\citenamefont{Heersche,
  Jarillo-Herrero, Oostinga, Vandersypen, and Morpurgo}}]{Heersche:nat07}
\bibinfo{author}{\bibfnamefont{H.~B.} \bibnamefont{Heersche}},
  \bibinfo{author}{\bibfnamefont{P.}~\bibnamefont{Jarillo-Herrero}},
  \bibinfo{author}{\bibfnamefont{J.~B.} \bibnamefont{Oostinga}},
  \bibinfo{author}{\bibfnamefont{L.~M.~K.} \bibnamefont{Vandersypen}},
  \bibnamefont{and} \bibinfo{author}{\bibfnamefont{A.~F.}
  \bibnamefont{Morpurgo}}, \bibinfo{journal}{Nature}
  \textbf{\bibinfo{volume}{446}}, \bibinfo{pages}{56} (\bibinfo{year}{2007}).

\bibitem[{\citenamefont{Novoselov et~al.}(2007)\citenamefont{Novoselov, Jiang,
  Zhang, Morozov, Stormer, Zeitler, Maan, Boebinger, Kim, and
  Geim}}]{Novoselov:sc07}
\bibinfo{author}{\bibfnamefont{K.~S.} \bibnamefont{Novoselov}},
  \bibinfo{author}{\bibfnamefont{Z.}~\bibnamefont{Jiang}},
  \bibinfo{author}{\bibfnamefont{Y.}~\bibnamefont{Zhang}},
  \bibinfo{author}{\bibfnamefont{S.~V.} \bibnamefont{Morozov}},
  \bibinfo{author}{\bibfnamefont{H.~L.} \bibnamefont{Stormer}},
  \bibinfo{author}{\bibfnamefont{U.}~\bibnamefont{Zeitler}},
  \bibinfo{author}{\bibfnamefont{J.~C.} \bibnamefont{Maan}},
  \bibinfo{author}{\bibfnamefont{G.~S.} \bibnamefont{Boebinger}},
  \bibinfo{author}{\bibfnamefont{P.}~\bibnamefont{Kim}}, \bibnamefont{and}
  \bibinfo{author}{\bibfnamefont{A.~K.} \bibnamefont{Geim}},
  \bibinfo{journal}{Science} \textbf{\bibinfo{volume}{315}},
  \bibinfo{pages}{1379} (\bibinfo{year}{2007}).

\bibitem[{\citenamefont{Hill et~al.}(2006)\citenamefont{Hill, Geim, Novoselov,
  Schedin, and Blake}}]{Hill:ieeem06}
\bibinfo{author}{\bibfnamefont{E.~W.} \bibnamefont{Hill}},
  \bibinfo{author}{\bibfnamefont{A.~K.} \bibnamefont{Geim}},
  \bibinfo{author}{\bibfnamefont{K.}~\bibnamefont{Novoselov}},
  \bibinfo{author}{\bibfnamefont{F.}~\bibnamefont{Schedin}}, \bibnamefont{and}
  \bibinfo{author}{\bibfnamefont{P.}~\bibnamefont{Blake}},
  \bibinfo{journal}{IEEE Trans. Mag.} \textbf{\bibinfo{volume}{42}},
  \bibinfo{pages}{2694} (\bibinfo{year}{2006}).

\bibitem[{\citenamefont{Tombros et~al.}(2007)\citenamefont{Tombros, Jozsa,
  Popinciuc, Jonkman, and van Wees}}]{Tombros:nat07}
\bibinfo{author}{\bibfnamefont{N.}~\bibnamefont{Tombros}},
  \bibinfo{author}{\bibfnamefont{C.}~\bibnamefont{Jozsa}},
  \bibinfo{author}{\bibfnamefont{M.}~\bibnamefont{Popinciuc}},
  \bibinfo{author}{\bibfnamefont{H.~T.} \bibnamefont{Jonkman}},
  \bibnamefont{and} \bibinfo{author}{\bibfnamefont{B.~J.} \bibnamefont{van
  Wees}}, \bibinfo{journal}{Nature} \textbf{\bibinfo{volume}{448}},
  \bibinfo{pages}{571} (\bibinfo{year}{2007}).

\bibitem[{\citenamefont{Ibach and L{\"{u}}th}(1995)}]{Ibach:95}
\bibinfo{author}{\bibfnamefont{H.}~\bibnamefont{Ibach}} \bibnamefont{and}
  \bibinfo{author}{\bibfnamefont{H.}~\bibnamefont{L{\"{u}}th}},
  \emph{\bibinfo{title}{Solid-State Physics}}
  (\bibinfo{publisher}{Springer-Verlag}, \bibinfo{address}{Berlin, Heidelberg},
  \bibinfo{year}{1995}), \bibinfo{edition}{2nd} ed.

\bibitem[{\citenamefont{Dedkov et~al.}(2008)\citenamefont{Dedkov, Fonin, and
  Laubschat}}]{Dedkov:apl08}
\bibinfo{author}{\bibfnamefont{Y.~S.} \bibnamefont{Dedkov}},
  \bibinfo{author}{\bibfnamefont{M.}~\bibnamefont{Fonin}}, \bibnamefont{and}
  \bibinfo{author}{\bibfnamefont{C.}~\bibnamefont{Laubschat}},
  \bibinfo{journal}{Appl. Phys. Lett.} \textbf{\bibinfo{volume}{92}},
  \bibinfo{pages}{052506} (\bibinfo{year}{2008}).

\bibitem[{\citenamefont{Oshima and Nagashima}(1997)}]{Oshima:jpcm97}
\bibinfo{author}{\bibfnamefont{C.}~\bibnamefont{Oshima}} \bibnamefont{and}
  \bibinfo{author}{\bibfnamefont{A.}~\bibnamefont{Nagashima}},
  \bibinfo{journal}{J. Phys.: Condens. Matter.} \textbf{\bibinfo{volume}{9}},
  \bibinfo{pages}{1} (\bibinfo{year}{1997}).

\bibitem[{\citenamefont{Gamo et~al.}(1997)\citenamefont{Gamo, Nagashima,
  Wakabayashi, Terai, and Oshima}}]{Gamo:ss97}
\bibinfo{author}{\bibfnamefont{Y.}~\bibnamefont{Gamo}},
  \bibinfo{author}{\bibfnamefont{A.}~\bibnamefont{Nagashima}},
  \bibinfo{author}{\bibfnamefont{M.}~\bibnamefont{Wakabayashi}},
  \bibinfo{author}{\bibfnamefont{M.}~\bibnamefont{Terai}}, \bibnamefont{and}
  \bibinfo{author}{\bibfnamefont{C.}~\bibnamefont{Oshima}},
  \bibinfo{journal}{Surface Science} \textbf{\bibinfo{volume}{374}},
  \bibinfo{pages}{61} (\bibinfo{year}{1997}).

\bibitem[{\citenamefont{Andersen et~al.}(1985)\citenamefont{Andersen, Jepsen,
  and Gl{\"{o}}tzel}}]{Andersen:85}
\bibinfo{author}{\bibfnamefont{O.~K.} \bibnamefont{Andersen}},
  \bibinfo{author}{\bibfnamefont{O.}~\bibnamefont{Jepsen}}, \bibnamefont{and}
  \bibinfo{author}{\bibfnamefont{D.}~\bibnamefont{Gl{\"{o}}tzel}}, in
  \emph{\bibinfo{booktitle}{Highlights of Condensed Matter Theory}}, edited by
  \bibinfo{editor}{\bibfnamefont{F.}~\bibnamefont{Bassani}},
  \bibinfo{editor}{\bibfnamefont{F.}~\bibnamefont{Fumi}}, \bibnamefont{and}
  \bibinfo{editor}{\bibfnamefont{M.~P.} \bibnamefont{Tosi}}
  (\bibinfo{publisher}{North-Holland}, \bibinfo{address}{Amsterdam},
  \bibinfo{year}{1985}), International School of Physics `Enrico Fermi',
  Varenna, Italy,, pp. \bibinfo{pages}{59--176}.

\bibitem[{\citenamefont{Bl{\"{o}}chl}(1994)}]{Blochl:prb94b}
\bibinfo{author}{\bibfnamefont{P.~E.} \bibnamefont{Bl{\"{o}}chl}},
  \bibinfo{journal}{Phys. Rev. B} \textbf{\bibinfo{volume}{50}},
  \bibinfo{pages}{17953} (\bibinfo{year}{1994}).

\bibitem[{\citenamefont{Kresse and Joubert}(1999)}]{Kresse:prb99}
\bibinfo{author}{\bibfnamefont{G.}~\bibnamefont{Kresse}} \bibnamefont{and}
  \bibinfo{author}{\bibfnamefont{D.}~\bibnamefont{Joubert}},
  \bibinfo{journal}{Phys. Rev. B} \textbf{\bibinfo{volume}{59}},
  \bibinfo{pages}{1758} (\bibinfo{year}{1999}).

\bibitem[{\citenamefont{Kresse and Hafner}(1993)}]{Kresse:prb93}
\bibinfo{author}{\bibfnamefont{G.}~\bibnamefont{Kresse}} \bibnamefont{and}
  \bibinfo{author}{\bibfnamefont{J.}~\bibnamefont{Hafner}},
  \bibinfo{journal}{Phys. Rev. B} \textbf{\bibinfo{volume}{47}},
  \bibinfo{pages}{558} (\bibinfo{year}{1993}).

\bibitem[{\citenamefont{Kresse and Furthmuller}(1996)}]{Kresse:prb96}
\bibinfo{author}{\bibfnamefont{G.}~\bibnamefont{Kresse}} \bibnamefont{and}
  \bibinfo{author}{\bibfnamefont{J.}~\bibnamefont{Furthmuller}},
  \bibinfo{journal}{Phys. Rev. B} \textbf{\bibinfo{volume}{54}},
  \bibinfo{pages}{11169} (\bibinfo{year}{1996}).

\bibitem[{\citenamefont{Neugebauer and Scheffler}(1992)}]{Neugebauer:prb92}
\bibinfo{author}{\bibfnamefont{J.}~\bibnamefont{Neugebauer}} \bibnamefont{and}
  \bibinfo{author}{\bibfnamefont{M.}~\bibnamefont{Scheffler}},
  \bibinfo{journal}{Phys. Rev. B} \textbf{\bibinfo{volume}{46}},
  \bibinfo{pages}{16067} (\bibinfo{year}{1992}).

\bibitem[{\citenamefont{Bl{\"{o}}chl et~al.}(1994)\citenamefont{Bl{\"{o}}chl,
  Jepsen, and Andersen}}]{Blochl:prb94a}
\bibinfo{author}{\bibfnamefont{P.~E.} \bibnamefont{Bl{\"{o}}chl}},
  \bibinfo{author}{\bibfnamefont{O.}~\bibnamefont{Jepsen}}, \bibnamefont{and}
  \bibinfo{author}{\bibfnamefont{O.~K.} \bibnamefont{Andersen}},
  \bibinfo{journal}{Phys. Rev. B} \textbf{\bibinfo{volume}{49}},
  \bibinfo{pages}{16223} (\bibinfo{year}{1994}).

\bibitem[{\citenamefont{Giovannetti et~al.}(2008)\citenamefont{Giovannetti,
  Khomyakov, Brocks, Karpan, van~den Brink, and Kelly}}]{Giovannetti:prl08}
\bibinfo{author}{\bibfnamefont{G.}~\bibnamefont{Giovannetti}},
  \bibinfo{author}{\bibfnamefont{P.~A.} \bibnamefont{Khomyakov}},
  \bibinfo{author}{\bibfnamefont{G.}~\bibnamefont{Brocks}},
  \bibinfo{author}{\bibfnamefont{V.~M.} \bibnamefont{Karpan}},
  \bibinfo{author}{\bibfnamefont{J.}~\bibnamefont{van~den Brink}},
  \bibnamefont{and} \bibinfo{author}{\bibfnamefont{P.~J.} \bibnamefont{Kelly}},
  \bibinfo{journal}{Phys. Rev. Lett.} \textbf{\bibinfo{volume}{101}},
  \bibinfo{pages}{026803} (\bibinfo{year}{2008}).

\bibitem[{\citenamefont{Ando}(1991)}]{Ando:prb91}
\bibinfo{author}{\bibfnamefont{T.}~\bibnamefont{Ando}}, \bibinfo{journal}{Phys.
  Rev. B} \textbf{\bibinfo{volume}{44}}, \bibinfo{pages}{8017}
  (\bibinfo{year}{1991}).

\bibitem[{\citenamefont{Khomyakov et~al.}(2005)\citenamefont{Khomyakov, Brocks,
  Karpan, Zwierzycki, and Kelly}}]{Khomyakov:prb05}
\bibinfo{author}{\bibfnamefont{P.~A.} \bibnamefont{Khomyakov}},
  \bibinfo{author}{\bibfnamefont{G.}~\bibnamefont{Brocks}},
  \bibinfo{author}{\bibfnamefont{V.}~\bibnamefont{Karpan}},
  \bibinfo{author}{\bibfnamefont{M.}~\bibnamefont{Zwierzycki}},
  \bibnamefont{and} \bibinfo{author}{\bibfnamefont{P.~J.} \bibnamefont{Kelly}},
  \bibinfo{journal}{Phys. Rev. B} \textbf{\bibinfo{volume}{72}},
  \bibinfo{pages}{035450} (\bibinfo{year}{2005}).

\bibitem[{\citenamefont{Xia et~al.}(2001)\citenamefont{Xia, Kelly, Bauer,
  Turek, Kudrnovsk\'{y}, and Drchal}}]{Xia:prb01}
\bibinfo{author}{\bibfnamefont{K.}~\bibnamefont{Xia}},
  \bibinfo{author}{\bibfnamefont{P.~J.} \bibnamefont{Kelly}},
  \bibinfo{author}{\bibfnamefont{G.~E.~W.} \bibnamefont{Bauer}},
  \bibinfo{author}{\bibfnamefont{I.}~\bibnamefont{Turek}},
  \bibinfo{author}{\bibfnamefont{J.}~\bibnamefont{Kudrnovsk\'{y}}},
  \bibnamefont{and} \bibinfo{author}{\bibfnamefont{V.}~\bibnamefont{Drchal}},
  \bibinfo{journal}{Phys. Rev. B} \textbf{\bibinfo{volume}{63}},
  \bibinfo{pages}{064407} (\bibinfo{year}{2001}).

\bibitem[{\citenamefont{Xia et~al.}(2006)\citenamefont{Xia, Zwierzycki,
  Talanana, Kelly, and Bauer}}]{Xia:prb06}
\bibinfo{author}{\bibfnamefont{K.}~\bibnamefont{Xia}},
  \bibinfo{author}{\bibfnamefont{M.}~\bibnamefont{Zwierzycki}},
  \bibinfo{author}{\bibfnamefont{M.}~\bibnamefont{Talanana}},
  \bibinfo{author}{\bibfnamefont{P.~J.} \bibnamefont{Kelly}}, \bibnamefont{and}
  \bibinfo{author}{\bibfnamefont{G.~E.~W.} \bibnamefont{Bauer}},
  \bibinfo{journal}{Phys. Rev. B} \textbf{\bibinfo{volume}{73}},
  \bibinfo{pages}{064420} (\bibinfo{year}{2006}).

\bibitem[{\citenamefont{Zwierzycki et~al.}(2008)\citenamefont{Zwierzycki,
  Khomyakov, Starikov, Xia, Talanana, Xu, Karpan, Marushchenko, Turek, Bauer
  et~al.}}]{Zwierzycki:pssb08}
\bibinfo{author}{\bibfnamefont{M.}~\bibnamefont{Zwierzycki}},
  \bibinfo{author}{\bibfnamefont{P.~A.} \bibnamefont{Khomyakov}},
  \bibinfo{author}{\bibfnamefont{A.~A.} \bibnamefont{Starikov}},
  \bibinfo{author}{\bibfnamefont{K.}~\bibnamefont{Xia}},
  \bibinfo{author}{\bibfnamefont{M.}~\bibnamefont{Talanana}},
  \bibinfo{author}{\bibfnamefont{P.~X.} \bibnamefont{Xu}},
  \bibinfo{author}{\bibfnamefont{V.~M.} \bibnamefont{Karpan}},
  \bibinfo{author}{\bibfnamefont{I.}~\bibnamefont{Marushchenko}},
  \bibinfo{author}{\bibfnamefont{I.}~\bibnamefont{Turek}},
  \bibinfo{author}{\bibfnamefont{G.~E.~W.} \bibnamefont{Bauer}},
  \bibnamefont{et~al.}, \bibinfo{journal}{phys. stat. sol. B}
  \textbf{\bibinfo{volume}{245}}, \bibinfo{pages}{623} (\bibinfo{year}{2008}).

\bibitem[{\citenamefont{B{\"{u}}ttiker
  et~al.}(1985)\citenamefont{B{\"{u}}ttiker, Imry, Landauer, and
  Pinhas}}]{Buttiker:prb85}
\bibinfo{author}{\bibfnamefont{M.}~\bibnamefont{B{\"{u}}ttiker}},
  \bibinfo{author}{\bibfnamefont{Y.}~\bibnamefont{Imry}},
  \bibinfo{author}{\bibfnamefont{R.}~\bibnamefont{Landauer}}, \bibnamefont{and}
  \bibinfo{author}{\bibfnamefont{S.}~\bibnamefont{Pinhas}},
  \bibinfo{journal}{Phys. Rev. B} \textbf{\bibinfo{volume}{31}},
  \bibinfo{pages}{6207} (\bibinfo{year}{1985}).

\bibitem[{\citenamefont{Datta}(1995)}]{Datta:95}
\bibinfo{author}{\bibfnamefont{S.}~\bibnamefont{Datta}},
  \emph{\bibinfo{title}{Electronic Transport in Mesoscopic Systems}}
  (\bibinfo{publisher}{Cambridge University Press},
  \bibinfo{address}{Cambridge}, \bibinfo{year}{1995}).

\bibitem[{\citenamefont{Turek et~al.}(1997)\citenamefont{Turek, Drchal,
  Kudrnovsk\'{y}, \v{S}ob, and Weinberger}}]{Turek:97}
\bibinfo{author}{\bibfnamefont{I.}~\bibnamefont{Turek}},
  \bibinfo{author}{\bibfnamefont{V.}~\bibnamefont{Drchal}},
  \bibinfo{author}{\bibfnamefont{J.}~\bibnamefont{Kudrnovsk\'{y}}},
  \bibinfo{author}{\bibfnamefont{M.}~\bibnamefont{\v{S}ob}}, \bibnamefont{and}
  \bibinfo{author}{\bibfnamefont{P.}~\bibnamefont{Weinberger}},
  \emph{\bibinfo{title}{Electronic Structure of Disordered Alloys, Surfaces and
  Interfaces}} (\bibinfo{publisher}{Kluwer},
  \bibinfo{address}{Boston-London-Dordrecht}, \bibinfo{year}{1997}).

\bibitem[{\citenamefont{Soven}(1967)}]{Soven:pr67}
\bibinfo{author}{\bibfnamefont{P.}~\bibnamefont{Soven}},
  \bibinfo{journal}{Phys. Rev.} \textbf{\bibinfo{volume}{156}},
  \bibinfo{pages}{809} (\bibinfo{year}{1967}).

\bibitem[{\citenamefont{Gl{\"{o}}tzel et~al.}(1980)\citenamefont{Gl{\"{o}}tzel,
  Segall, and Andersen}}]{Glotzel:ssc80}
\bibinfo{author}{\bibfnamefont{D.}~\bibnamefont{Gl{\"{o}}tzel}},
  \bibinfo{author}{\bibfnamefont{B.}~\bibnamefont{Segall}}, \bibnamefont{and}
  \bibinfo{author}{\bibfnamefont{O.~K.} \bibnamefont{Andersen}},
  \bibinfo{journal}{Sol. State Comm.} \textbf{\bibinfo{volume}{36}},
  \bibinfo{pages}{403} (\bibinfo{year}{1980}).

\bibitem[{\citenamefont{Hahn}(2005)}]{ITC}
\bibinfo{editor}{\bibfnamefont{T.}~\bibnamefont{Hahn}}, ed.,
  \emph{\bibinfo{title}{International Tables for Crystallography}},
  vol.~\bibinfo{volume}{A} (\bibinfo{publisher}{Springer},
  \bibinfo{year}{2005}).

\bibitem[{\citenamefont{Bertoni et~al.}(2005)\citenamefont{Bertoni, Calmels,
  Altibelli, and Serin}}]{Bertoni:prb05}
\bibinfo{author}{\bibfnamefont{G.}~\bibnamefont{Bertoni}},
  \bibinfo{author}{\bibfnamefont{L.}~\bibnamefont{Calmels}},
  \bibinfo{author}{\bibfnamefont{A.}~\bibnamefont{Altibelli}},
  \bibnamefont{and} \bibinfo{author}{\bibfnamefont{V.}~\bibnamefont{Serin}},
  \bibinfo{journal}{Phys. Rev. B} \textbf{\bibinfo{volume}{71}},
  \bibinfo{pages}{075402} (\bibinfo{year}{2005}).

\bibitem[{not()}]{note3}
\bibinfo{note}{
  To describe the adsorption of graphite on Co and Ni within model I we use two ESs
  with a WSR $r =1.12$ a.u.~ at B and C sites, positioned $1.57$ a.u. above
  the metal surface. For model II we use one ES with a WSR $r = 1.7$ a.u. at a C
  site, at $2.42$ a.u. above the surface. If the distance between graphite and
  the metal surface is larger, which is the case for graphite on Cu, we use three
  ESs with  WSR $r = 1.6$ a.u., at A, B, and C sites, respectively, $3.33$ a.u.
  above the surface.}

\bibitem[{\citenamefont{Butler et~al.}(2001)\citenamefont{Butler, Zhang,
  Schulthess, and MacLaren}}]{Butler:prb01}
\bibinfo{author}{\bibfnamefont{W.~H.} \bibnamefont{Butler}},
  \bibinfo{author}{\bibfnamefont{X.-G.} \bibnamefont{Zhang}},
  \bibinfo{author}{\bibfnamefont{T.~C.} \bibnamefont{Schulthess}},
  \bibnamefont{and} \bibinfo{author}{\bibfnamefont{J.~M.}
  \bibnamefont{MacLaren}}, \bibinfo{journal}{Phys. Rev. B}
  \textbf{\bibinfo{volume}{63}}, \bibinfo{pages}{054416}
  (\bibinfo{year}{2001}).

\bibitem[{\citenamefont{Mathon and Umerski}(2001)}]{Mathon:prb01}
\bibinfo{author}{\bibfnamefont{J.}~\bibnamefont{Mathon}} \bibnamefont{and}
  \bibinfo{author}{\bibfnamefont{A.}~\bibnamefont{Umerski}},
  \bibinfo{journal}{Phys. Rev. B} \textbf{\bibinfo{volume}{63}},
  \bibinfo{pages}{220403(R)} (\bibinfo{year}{2001}).

\bibitem[{\citenamefont{Charlier et~al.}(1991)\citenamefont{Charlier, Gonze,
  and Michenaud}}]{Charlier:prb91a}
\bibinfo{author}{\bibfnamefont{J.-C.} \bibnamefont{Charlier}},
  \bibinfo{author}{\bibfnamefont{X.}~\bibnamefont{Gonze}}, \bibnamefont{and}
  \bibinfo{author}{\bibfnamefont{J.-P.} \bibnamefont{Michenaud}},
  \bibinfo{journal}{Phys. Rev. B} \textbf{\bibinfo{volume}{43}},
  \bibinfo{pages}{4579} (\bibinfo{year}{1991}).

\bibitem[{\citenamefont{Giovannetti et~al.}(2007)\citenamefont{Giovannetti,
  Khomyakov, Brocks, Kelly, and van~den Brink}}]{Giovannetti:prb07}
\bibinfo{author}{\bibfnamefont{G.}~\bibnamefont{Giovannetti}},
  \bibinfo{author}{\bibfnamefont{P.~A.} \bibnamefont{Khomyakov}},
  \bibinfo{author}{\bibfnamefont{G.}~\bibnamefont{Brocks}},
  \bibinfo{author}{\bibfnamefont{P.~J.} \bibnamefont{Kelly}}, \bibnamefont{and}
  \bibinfo{author}{\bibfnamefont{J.}~\bibnamefont{van~den Brink}},
  \bibinfo{journal}{Phys. Rev. B} \textbf{\bibinfo{volume}{76}},
  \bibinfo{pages}{073103} (\bibinfo{year}{2007}).

\bibitem[{\citenamefont{Geim and Novoselov}(2007)}]{Geim:natm07}
\bibinfo{author}{\bibfnamefont{A.~K.} \bibnamefont{Geim}} \bibnamefont{and}
  \bibinfo{author}{\bibfnamefont{K.~S.} \bibnamefont{Novoselov}},
  \bibinfo{journal}{Nature Materials} \textbf{\bibinfo{volume}{6}},
  \bibinfo{pages}{183} (\bibinfo{year}{2007}).

\bibitem[{\citenamefont{Monsma et~al.}(1998)\citenamefont{Monsma, Vlutters, and
  Lodder}}]{Monsma:sc98}
\bibinfo{author}{\bibfnamefont{D.~J.} \bibnamefont{Monsma}},
  \bibinfo{author}{\bibfnamefont{R.}~\bibnamefont{Vlutters}}, \bibnamefont{and}
  \bibinfo{author}{\bibfnamefont{J.~C.} \bibnamefont{Lodder}},
  \bibinfo{journal}{Science} \textbf{\bibinfo{volume}{281}},
  \bibinfo{pages}{407} (\bibinfo{year}{1998}).

\bibitem[{\citenamefont{Matsubara et~al.}(1990)\citenamefont{Matsubara,
  Sugihara, and Tsuzuku}}]{Matsubara:prb90}
\bibinfo{author}{\bibfnamefont{K.}~\bibnamefont{Matsubara}},
  \bibinfo{author}{\bibfnamefont{K.}~\bibnamefont{Sugihara}}, \bibnamefont{and}
  \bibinfo{author}{\bibfnamefont{T.}~\bibnamefont{Tsuzuku}},
  \bibinfo{journal}{Phys. Rev. B} \textbf{\bibinfo{volume}{41}},
  \bibinfo{pages}{969} (\bibinfo{year}{1990}).

\bibitem[{\citenamefont{Schmidt et~al.}(2000)\citenamefont{Schmidt, Ferrand,
  Molenkamp, Filip, and van Wees}}]{Schmidt:prb00}
\bibinfo{author}{\bibfnamefont{G.}~\bibnamefont{Schmidt}},
  \bibinfo{author}{\bibfnamefont{D.}~\bibnamefont{Ferrand}},
  \bibinfo{author}{\bibfnamefont{L.~W.} \bibnamefont{Molenkamp}},
  \bibinfo{author}{\bibfnamefont{A.~T.} \bibnamefont{Filip}}, \bibnamefont{and}
  \bibinfo{author}{\bibfnamefont{B.~J.} \bibnamefont{van Wees}},
  \bibinfo{journal}{Phys. Rev. B} \textbf{\bibinfo{volume}{62}},
  \bibinfo{pages}{R4790} (\bibinfo{year}{2000}).

\end{thebibliography}

\end{document}